\documentclass[twocolumn,english,floatfix]{revtex4}
\usepackage[T1]{fontenc}
\usepackage[latin9]{inputenc}
\usepackage{color}
\usepackage{soul}
\usepackage{multirow}
\usepackage{graphicx}
\usepackage{calc}
\usepackage{amsmath}
\usepackage{babel}
\usepackage{float}
\usepackage{url} 

\usepackage[paperwidth=210mm,paperheight=297mm,centering,hmargin=2cm,vmargin=2.5cm]{geometry}

\usepackage{xargs}                     
\usepackage[pdftex,dvipsnames]{xcolor} 
\makeatletter

\providecommand{\tabularnewline}{\\}

\usepackage{babel}
\makeatother

\begin{document}
\title{Bright \AA ngstrom and picometer free electron laser based on the Large Hadron electron Collider energy recovery linac}
\author{Z.~Nergiz}\email{znergiz@ohu.edu.tr}
\affiliation{Nigde Omer Halisdemir University, Faculty of Arts and Sciences, Physics
Department, 51200 Nigde, Turkey}

\author{N.S.~Mirian}
 \email{najmeh.mirian@desy.de}
\affiliation{ Deutsches Elektronen-Synchrotron (DESY), 22607 Hamburg, Germany; \\ and European Organization for Nuclear Research (CERN), 1211 Geneva 23, Switzerland
}

\author{A.~Aksoy}
\affiliation{Ankara University, Institute of Accelerator Technologies, 06830, Ankara,
Turkey; \\  and Turkish Accelerator and Radiation Laboratory, 06830, Ankara,
Turkey}

\author{D.~Zhou}
\affiliation{High Energy Accelerator Research Organization (KEK), Tsukuba, Ibaraki, Japan}

\author{F.~Zimmermann}
\affiliation{European Organization for Nuclear Research (CERN), 1211 Geneva 23,
Switzerland}

\author{H.~Aksakal}
\affiliation{Kahraman Maras Sutcu Imam University, 46040 Kahramanmaras, Turkey}
%%%%%
\begin{abstract}
The Large Hadron electron Collider (LHeC) is a proposed future
particle-physics project colliding 60 GeV electrons from a
six-pass  
recirculating energy-recovery linac (ERL) with 7 TeV protons stored in
the LHC. The ERL technology allows for much higher beam current and,
therefore, higher luminosity than a traditional linac. The
high-current, high-energy electron beam can also be used to drive a
free electron laser (FEL). In this study, we investigate the
performance of an LHeC-based FEL, operated in the self-amplified
spontaneous emission mode using electron beams 
after one or two turns, 
with beam energies of, e.g., 10, 20, 30 and 40 GeV,
and aim at producing X-ray pulses at wavelengths 
ranging from 8~\AA\ to 0.5~\AA . 
In addition, we explore a possible path to use the 40 GeV
electron beam for generating photon pulses  
at much lower wavelengths, down to a few picometer. 
We demonstrate that such ERL-based high-energy  
FEL would have the potential to provide orders
of magnitude higher average brilliance at \AA\ wavelengths
than any  other FEL either existing or proposed.
It might also allow a pioneering step 
into the picometer wavelength regime. 
\end{abstract}
\pacs{Valid PACS appear here}
\maketitle
%%$%%%%%%%%%%%%%%%%%%%%%%$
\section{\label{sec:level1}Introduction}
The Large Hadron electron Collider (LHeC) \cite{LHeCdesign} 
 is a proposed future lepton-hadron collider at CERN, which would be realized 
by colliding protons circulating in one of the existing rings of the Large Hadron Collider (LHC) 
with a 60 GeV electron beam from a 
six-pass recirculating racetrack-shape energy-recovery linac (ERL). 
The electron beam consists of bunches 
of $3\times10^{9}$ particles each, spaced by 25 ns like the proton bunches, 
with an average beam current of about 20 mA \cite{lhechf}.
A recent design variant considers a lower electron 
beam energy of 50 GeV,
accompanied by a possibly higher beam current 
of up to 50 mA \cite{agostini2020lhec}.

The high-current ERL of the LHeC would also provide the opportunity for driving a Free Electron Laser (FEL) \cite{schopper}. Indeed, ERL-based FELs already operated, and operate, successfully in the electron-energy range of 10 to 200 MeV, e.g.~at BINP \cite{binp}, JAEA \cite{jaea} and JLAB \cite{jlab}. Their parameters are compiled in Table \ref{tab1}. 
A superconducting ERL with a higher beam energy of 0.5--1.0 GeV 
was proposed to produce 13.5 nm radiation, at 5 kW average power \cite{vinokurov2}.  
Another proposal with 5 GeV beam energy aimed at generating  X-rays at 
\AA\ wavelengths \cite{Bilderback_2010}. 
All of these operating or proposed facilities featured, or feature, 
a significantly lower 
beam energy than the LHeC FEL. 
Most similar to the LHeC-based FEL would be a possible upgrade of the European XFEL also based on an ERL-type of operation, with 100\% duty factor and an average brightness of $1.64\times10^{25}$ photons/s/mm$^{2}$/mrad$^{2}$/0.1\%
bandwidth at 8.5 GeV beam energy \cite{sekutowicz}. 

\begin{table}
\caption{Parameters of some operating ERL-based FELs.}
\label{tab1} \centering %
\begin{tabular}{lccc}
\hline 
Facility  & BINP  & JAEA  & JLAB
\tabularnewline
\hline 
% status &  active & active & past & proposed \\
Beam energy [MeV]  & 20  & 17  & 120\\
Peak current [A]  & 3000  & 35  & 300\\
Average current [mA]  & 100  & 8  & 8\\
Photon wavelength [$\mu$m]  & 40  & 22  & 1.6\\
Average FEL power [W] & 500  & 1  & 10,000\\
Pulse duration [ps]  & 50  & 0.32  & 0.17\\
\hline 
% peak brilliance [B] &  &  & \\
\end{tabular}
\end{table}

Though the LHeC is designed for energy frontier electron-hadron scattering experiments at the LHC, it is conceivable that the ERL program can be temporarily redefined, independently of electron-hadron operation, as, for example, during the decade in which the LHC may possibly be reconfigured to double its hadron beam energy within the High Energy LHC (HE-LHC) proposal~\cite{HELHC}, 
and during which no lepton-hadron collisions would take place.

In view of the performance expected from the LHeC-FEL (see Section \ref{sec:brilliance}) 
also the construction of a dedicated ERL-based X-ray FEL user facility could, 
and perhaps should, be considered.

\section{\label{sec:level2lhec} Adapting the LH\lowercase{e}C}

The ERL of LHeC is of racetrack shape. 
For the proposed collider operation, 
a 500 MeV electron bunch coming from the 
injector would be 
accelerated in each of two 10 GV 
superconducting linacs during three revolutions, after which it has obtained an energy of 60 GeV. 
Three additional revolutions, now with deceleration instead of acceleration, reconvert 
the energy stored in the beam back to radiofrequency (RF) energy~\cite{LHeCdesign}. The beam emittance 
and the energy spread of the particle beam increase with beam energy due to quantum fluctuations.

\begin{figure}[htbp]
\includegraphics[width=0.45\textwidth]{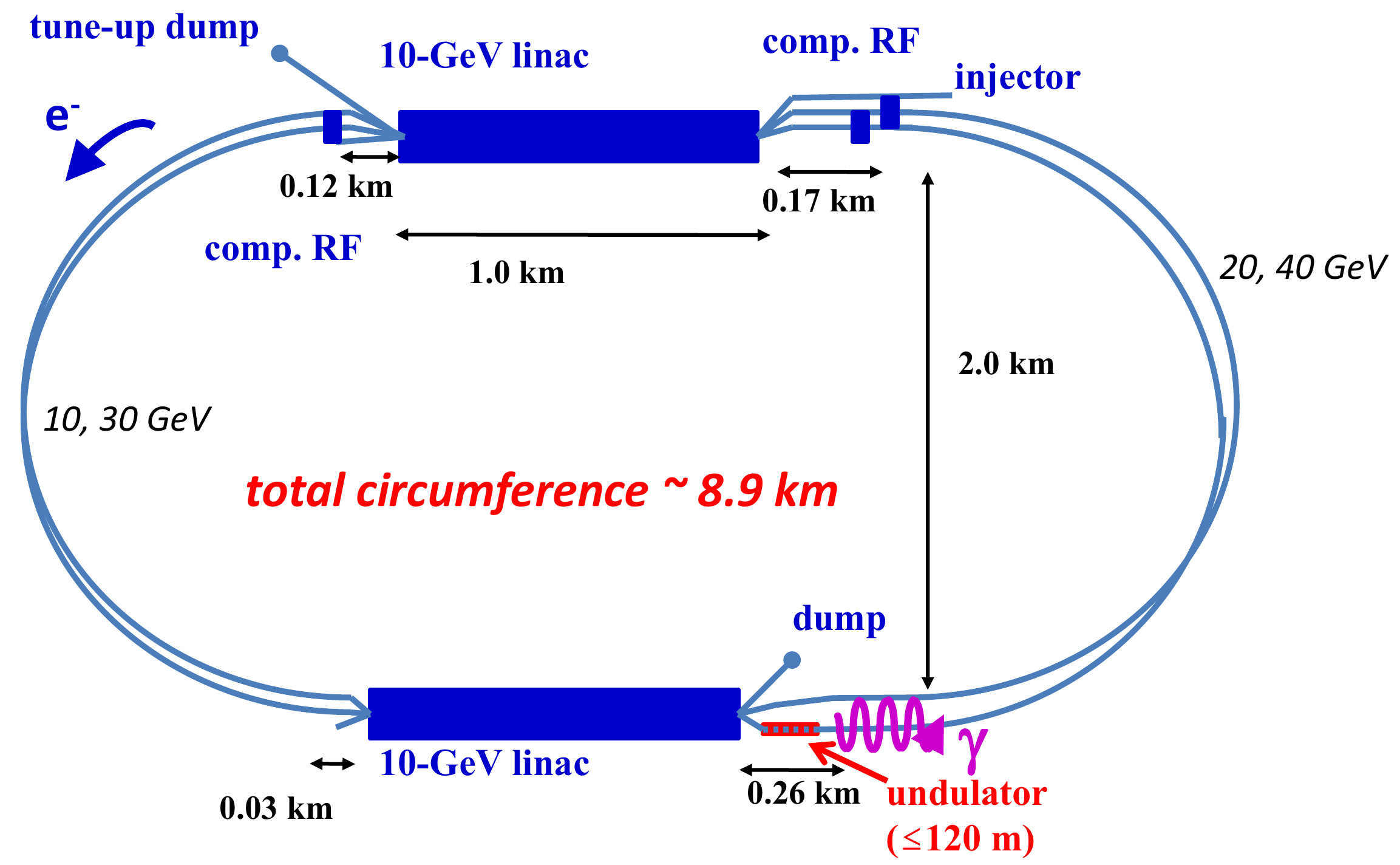}
\caption{\label{fig:scheme} LHeC recirculating linac reconfigured for FEL operation.}
\end{figure}

For the LHeC proper, the electron-beam emittance is not critical, since the proton-beam 
emittance is quite large.
Incoherent synchrotron radiation significantly increases 
the normalized rms 
emittance during the arc passages at 40 and 50 GeV beam energy,
by about 7~$\mu$m \cite[Table 7.14]{LHeCdesign}.
 However, in order to obtain coherent 
 X-rays at low wavelengths in FEL operation the beam emittance must be sufficiently small. 
Partly because of this emittance requirement, for the FEL operation,  
we choose the electron beam energy as 40 GeV  or lower, 
depending on the X-ray wavelength desired, rather than 60 GeV.
Figure \ref{fig:scheme} illustrates the LHeC ERL-FEL configuration.

The beam energy of 40 GeV can be attained after two passes through 
the two 10 GeV linacs, instead of the three passes of the standard LHeC operation. 
The subsequent deceleration would also happen 
during two additional passes. An energy of 20 GeV 
would already be achieved after a single pass through the two linacs, 
again followed by another pass of deceleration.
Beam energies of 10 and 30 GeV are also readily obtained after one or two turns,
with appropriate linac voltages and phasing. 

At high beam energy, the incoherent synchrotron radiation in the arcs 
blows up the energy spread and the transverse emittance. 
At a beam energy of 40 GeV, the accumulated relative energy spread induced by quantum 
fluctuations in the third LHeC arc is $5.3\times10^{-5}$
(2 MeV) \cite[Table 7.13]{LHeCdesign}. 
By contrast, at 20 GeV the additional 
energy spread due to 
incoherent synchrotron radiation (ISR) is  
negligible. For the chosen optics, the minimum additional 
contribution to the normalized emittance from incoherent 
synchrotron radiation is about 0.5~$\mu$m at 40 GeV 
\cite[Table 7.14]{LHeCdesign}, 
which is to be added to the initial emittance. 
At 20 GeV, the ISR effect, also 
on the transverse emittance, can be neglected.
Instead, here, the transverse normalized emittance may be limited solely 
by the performance of the RF gun, and the total emittance could be as low as, or lower than, 
0.5~$\mu$m for a bunch charge of 0.5 nC (so-called PITZ scaling)  \cite{bcarlsten2013}. 

%%%%%%%%%%%%%%%%%%%%%%%%%%%%%%%%%%%%%%%%%%%
%%%%%%%%%%%%%%%%%%%%%%%%%%%%%%%%%%%%%%%%%%
\section{\label{sec:comp}Bunch Compression Concept}

Compared with chicane- or wiggler-based bunch compression in a single-pass
linac \cite{kheifets,tor,trfz}, a recirculating linac offers additional
degrees of freedom to compress the bunch and also to tailor its longitudinal
profile, respectively, e.g.~by exploiting the linear momentum compaction
in the return arcs of the recirculating linac, and adjusting the RF
phases for each linac pass. Additional manipulations would be possible
by controlling (and cancelling) the second-order momentum compaction
through arc sextupole magnets \cite{kamitani}. For example, choosing
the proper linac configuration, in the downstream SLC (SLAC Linear
Collider) arcs the rms bunch length could be compressed by more than
an order of magnitude, from above 1~mm down to about 50~$\mu$m
\cite{slc}.

To examine the possible LHeC ERL bunch-length compression for FEL
operation, we accelerate the beam off-crest in some of the first three
linac passages, and exploit the momentum dependent path length for
the first three LHeC ERL arcs \cite{bogaczarc,LHeCdesign}, which
for the CDR optics, including spreaders and combiners, amount to $R_{56}^{(1)}=R_{56}^{(2)}=0.21$~m
and $R_{56}^{(3)}=-0.31$~m, where the superindex in parentheses
counts the arc, and a positive value for the fifth coordinate, $z>0$,
refers to a particle ahead of the synchronous particle.

However, the additional energy spread due to incoherent synchrotron
radiation induced in the third arc $\Delta\sigma_{\delta}^{(3)}\approx5\times10^{-5}$,
along with the rather large (absolute) design value of $|R_{56}^{(3)}|=0.31$~m,
contributes to the final bunch length a minimum amount 
of $\Delta\sigma_{z,{\rm min}}\ge|R_{56}^{(3})|\Delta\sigma_{\delta}^{(3)}/\sqrt{3}\approx9$~$\mu$m,
not yet including any nonlinear contributions. Hence, this optics
does not allow squeezing the rms bunch length to values much below 10~$\mu$m.

In view of this limit, and 
profiting from the flexible momentum compaction (FMC)
arc optics,
we have explored the possibility of changing the optics of arc 3, 
so as to be similar to
those of arcs 1 and 2, % By optimizing the arc 3 quadrupole strenghts,
or even further reducing the (absolute) value of $R_{56}^{(3)}$.
% to $0.15$~m, thereby, 
allowing the compression to significantly shorter bunch lengths.
The possibility to compress to shorter bunches, however, comes at
the expense of a larger $I_{5}$ radiation integral.
Synchrotron radiation
in arc 3 then increases the horizontal normalized emittance 
to total values well above 2~$\mu$m.
This emittance would be too large for the 
FEL wavelengths we are targeting.

We have, therefore, proceeded with the arc-3 optics from the LHeC CDR,
which limits the possible compression to final rms bunch lengths 
not much below 10 $\mu$m, but provides for a smaller transverse emittance,
below 1~$\mu$m. 

In addition to the incoherent synchrotron radiation, also the effects
of wake fields and coherent synchrotron radiation need to be taken
into account.

\section{Shielded Coherent Radiation}

The large bending radius of the LHeC, $\rho\approx750$~m, combined
with a small vacuum chamber, suppresses the emission of synchrotron
radiation at long wavelengths and, in particular, the emission of
CSR \cite{schwinger,nodvick}. Specifically, synchrotron radiation
is shielded at wavelengths longer than \cite{laslett,warnockm,warnock}\cite[Eq.~(178)]{PhysRevSTAB.12.094402}\cite{zhoujjap}
\begin{equation}
\lambda_{{\rm sh}}\approx2\sqrt{\frac{d^{3}}{\rho}}\;,
\end{equation}
or, equivalently, for bunch lengths exceeding 
\begin{equation}
\sigma_{z,{\rm sh}}\approx\sqrt{\frac{d^{3}}{\rho\pi^{2}}}\;,
\end{equation}
where $d$ denotes the beam pipe diameter \cite{warnock}. Considering
the LHeC FEL, for $\rho\approx750$~m and $d\approx20$~mm, we find
$\sigma_{z,{\rm sh}}\approx30$~$\mu$m. With a reduced pipe diameter
of $d\approx 10$~mm, we would expect to obtain
complete CSR shielding down to $\sigma_{z,{\rm sh}}\le 12$~$\mu$m.

\begin{figure*}
\includegraphics[width=0.45\textwidth,height=5cm]{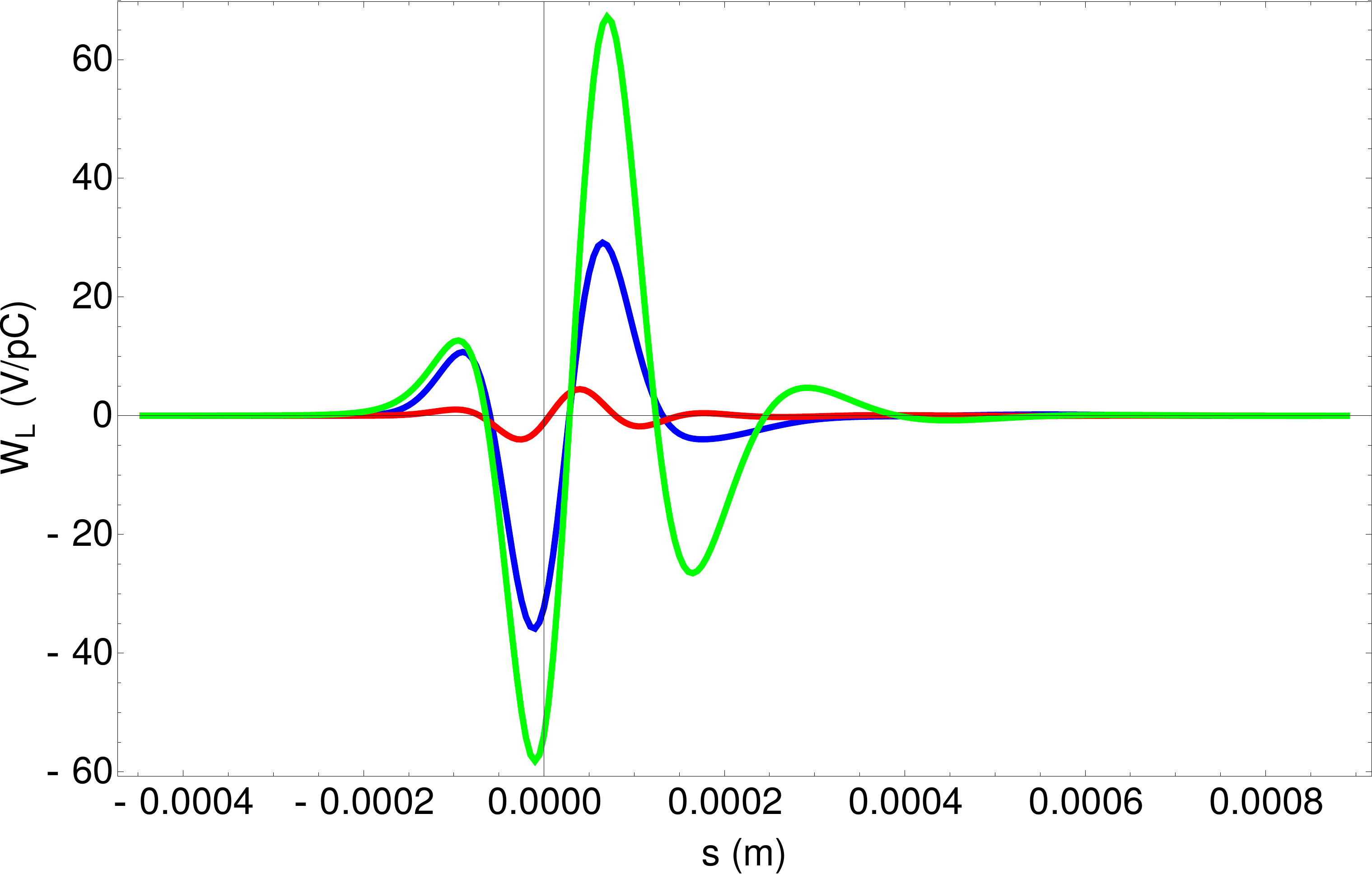} 
\includegraphics[width=0.45\textwidth,height=5cm]{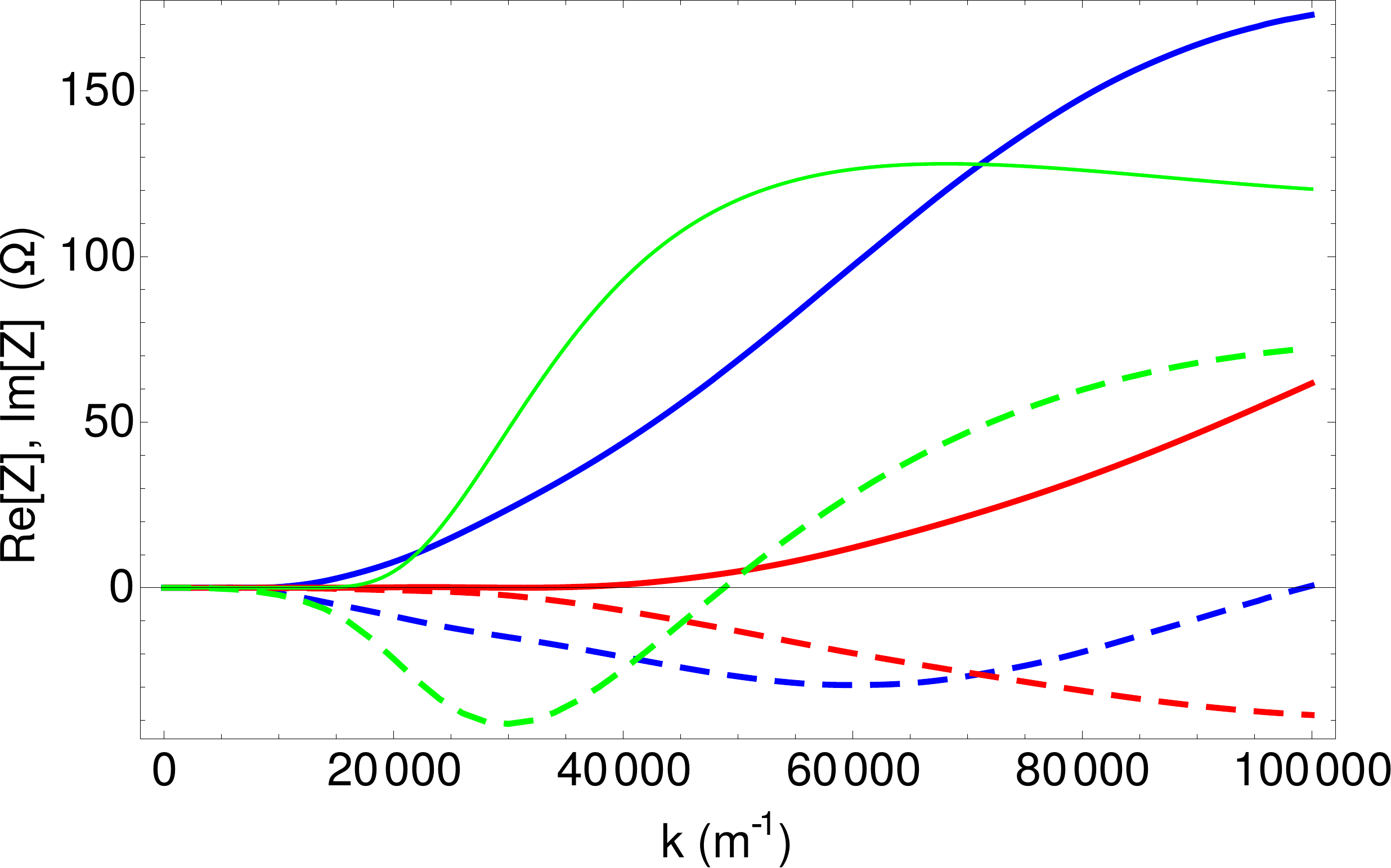}
\caption{\label{fig:csrw2}CSR wake field for a Gaussian bunch with 50 micron
bunch length (left) and impedance (right) for a 4-m long arc dipole
with $\rho=744$~m computed by the code CSRZ \cite{zhoujjap} for
a square beam pipe with 2 cm (blue) or 1 cm full aperture (red) in
the horizontal and vertical direction, compared with the CSR wake
and impedance calculated for a simple parallel plate model with a
vertical gap of 2 cm (green). In the right picture, solid lines refer
to the real part, dashed lines to the imaginary part of the impedance.}
\label{csrz3} 
\end{figure*}

A few programs are available to simulate the shielding for a realistic
closed vacuum chamber, rather than in free space or with parallel-plate boundaries. 
We employ the code CSRZ \cite{zhoujjap} to
compute the CSR impedance in the frequency domain for an LHeC arc dipole of
length 4 m, with a bending radius $\rho\approx750$~m. The shielding
calculation considers a square vacuum chamber with variable curvature of the beam orbit. The CSR wake field  
can be calculated from the impedance by convolution with the spectrum of a given longitudinal bunch profile \cite{zhouipac12}. 
Figure \ref{csrz3} compares the CSR impedance of an LHeC
arc dipole and the resulting wake function for a 50~$\mu$m long
bunch (blue curves) with those expected from a parallel-plate model,
for a full aperture $d$ of 2 cm (green curve). It also illustrates the
further dramatic reduction of the CSR impedance and wake field if
the square chamber size is reduced to 1 cm (red curve). The maximum
wave number $k_{{\rm max}}$ ($k=\omega/c=2\pi f/c$) corresponds
to about $5/\sigma_{z}$ with a typical 50~$\mu$m rms bunch length
in the third arc. Taking into account the bunch lengths in the different
arcs (see Table \ref{linacbp}) for the first arc we choose a cutoff 
wave number $k_{{\rm max}}$ of $30000$~m$^{-1}$,
for the second arc 60000~m$^{-1}$, and for the last arc 100000~m$^{-1}$. 
% The CSR impedance could further be reduced by decreasing the vertical
% gap to 1 cm. 
In the tracking simulations performed with the code ELEGANT, at each
dipole we include the CSR impedance, from CSRZ, corresponding to a
full vertical and horizontal chamber aperture of 1 cm.

We neglect the possible interference between CSR wake fields from
consecutive dipole magnets, but apply the CSR impedance independently
in each dipole magnet, which, from past experience for other accelerators,
represents a good first approximation.

We note that the suppression of both incoherent SR and CSR by the
vacuum chamber has been well proven experimentally. For the SLC collider
arcs, with $d=10$~mm and $\rho=280$~m, CSR should be shielded at
$\sigma_{z}>20$~$\mu$m, fully consistent with the complete absence
of any CSR effects in the observed beam evolution for minimum bunch
lengths around 50 $\mu$m \cite{slc}. A later series of dedicated
shielding studies at the BNL ATF further corroborated the theoretical
predictions for CSR shielding \cite{yakimenko}. Additional experimental
evidence for the suppression of (in this case, incoherent) synchrotron radiation
by the vacuum chamber, and for the predicted dependence on the bending
radius, comes from RHIC, where fully stripped gold ion (Au$^{+79}$)
experienced a nearly total suppression of synchrotron radiation (energy
loss per turn reduced by more than a factor ten) at an energy of 70
GeV/nucleon, and still a reduction by a factor larger than two at
100 GeV/nucleon \cite{abreu}. These experimental results are consistent
with CSRZ simulations.

\section{\label{sec:wake}Wake Fields}
The transverse and longitudinal wake fields in the LHeC 
linac RF cavities are
modeled using the short-range wake functions of \cite[Eq.~(2.17)]{Pellegrini:2235763}, which are 
based on Refs.~\cite{Bane:2003du,Calaga:2020926} (also see the 
illustration in \cite[Fig.~2.2]{Pellegrini:2235763}). 
More precisely, 
the simulation uses the wake potentials (Green function wake fields), and then computes the bunch wake field
from the actual particle distribution whenever the tracked bunch passes through a cavity.  
To illustrate the magnitude of the LHeC linac wake field, the nominal wake potential results in a longitudinal loss factor of 2.6 V/pC  per cavity in case of a Gaussian bunch with 2 mm rms length.

Resistive-wall wake fields may set a lower limit on the acceptable 
vacuum chamber dimension in the arcs. 
The material of the LHeC vacuum chamber has not been decided.
It could be made from copper or aluminum, and possibly be coated
\cite{LHeCdesign}. 
The characteristics of the resistive wall wake field is determined
by the parameter $s_{0}\equiv ( d^2 \rho_{\rm res}/(2 Z_{0}))^{1/3}$
\cite{Chao:1993zn,Bane:1995eu}, 
with $Z_{0}$ the impedance of free space (about 120$\pi$~$\Omega$).
The wake function is approximately constant over distances 
much shorter than $s_{0}$, but it oscillates over distances 
of a few $s_{0}$. 
Assuming that the LHeC arc vacuum chamber is made from copper, 
with a resistivity of $\rho_{\rm res}=1.7 \times 10^{-8}$~$\Omega$~m, 
for the smallest chamber aperture considered, $d=10$~mm, 
we obtain $s_{0}\approx 13$~$\mu$m, and the LHeC FEL 
bunches extend over several times $s_{0}$.
In this regime the average energy loss over a section of length $L$ 
is well approximated by \cite{Bane:1995eu,Piwinski:1994cp}
\begin{equation}
\Delta E _{\rm r.w.} \approx  - \frac{Z_{0}c L N_{b}e^2}{5 \pi d^2} \left(\frac{s_{0}}{\sigma_{z}}\right) ^{3/2}\; ,
\end{equation}
where $N_b$ denotes the bunch population, 
$\sigma_{z}$ the rms bunch length, and 
$L$ the length of the section in question (e.g.~$L\approx 3$~km for one arc).
Also using $d=10$~mm, $\sigma_{z}=100$~$\mu$m 
and $N_{b}=3\times 10^{9}$, we find for the 
average energy loss in the first arc   
$\Delta E _{\rm r.w.}\approx 4.9$~MeV.
The rms energy spread induced by the
resistive wall wake field will be of 
similar magnitude as the average energy loss. 
The estimated value of
$\Delta E_{\rm r.w.}$
is about twice as high as the average energy loss 
due to the RF cavity
wake fields in one linac.  

Another possible concern 
is the transverse
resistive wall wake field. 
The single-bunch resistive wall jitter amplification 
when passing through one arc, 
can be estimated, by 
averaging over several betatron oscillation periods,
as \cite{achao82,nlczdr}
\begin{equation}
G\equiv \beta_{y} \frac{\Delta y'}{\Delta y}
\approx \beta_{y} \pi^{2} \frac{N_{b}r_{e}}{\gamma \sigma_{z}}\frac{L}{d^{3}} \left< f_{R} \right> \sqrt{\lambda \sigma_{z}}\; ,
\label{gfac} 
\end{equation}
where $r_{e}$ denotes the classical 
electron radius, $L$ again the length of the section in question (e.g.~$L\approx 3$~km for one arc),
$\gamma$ the Lorentz factor, and $\lambda = \rho_{\rm res}/(120\pi\; \Omega)$.
At a beam energy of 10 GeV, assuming a copper beam pipe, 
and using $\left<f_{R}\right> =0.82$, $d=10$~mm,  
$\sigma_{z}=100$~$\mu$m, $N_{b}=3\times 10^{9}$ and $\beta_{x,y}
\approx 50$~m, we find
$G\approx 0.3$, which appears acceptable.

\section{\label{sec:comp-sim}Bunch Compression Simulations}

Realistic longitudinal tracking simulations of two full circulations
(four linac passages and three arc traversals) are performed with the code
ELEGANT \cite{elegant}, which can take into account not only the
linear and nonlinear optics, but, optionally, also the longitudinal
and transverse linac wake fields, incoherent synchrotron radiation,
and in addition, with an external ``impedance'' file, the effect of the shielded
coherent synchrotron radiation in the arc dipole magnets, 
as computed by CSRZ.  We have included all of these effects. 
However, our tracking simulations 
did not consider the (material-dependent)  
resistive wall wake field in the arcs. 
The CSR ``impedance'' file was varied   
according to the local bunch length. 
Below we present results for the nominal linac wake fields. 
We have also performed some 
simulations with a factor of 5 larger linac wake fields
(that is, five times larger wake potentials), 
yielding quite similar results.

For every case, 
we have optimized the RF phases in each linac 
to achieve highest peak
current after the third (first) arc or fourth (second)
linac passage, and adjusted the
linac RF voltage to maintain the target beam energy of 
10, 20, 30 or 40 GeV. 
As a result of this optimization process, at 40 GeV   
the RF voltage of linac 1 was reduced to 9.7 GV 
and the one of linac 2 raised to 11.4 GV, for all passages.

 %%%%%%%%%%%%%

\begin{figure}[htbp]
\includegraphics[width=0.48\textwidth,height=3cm]{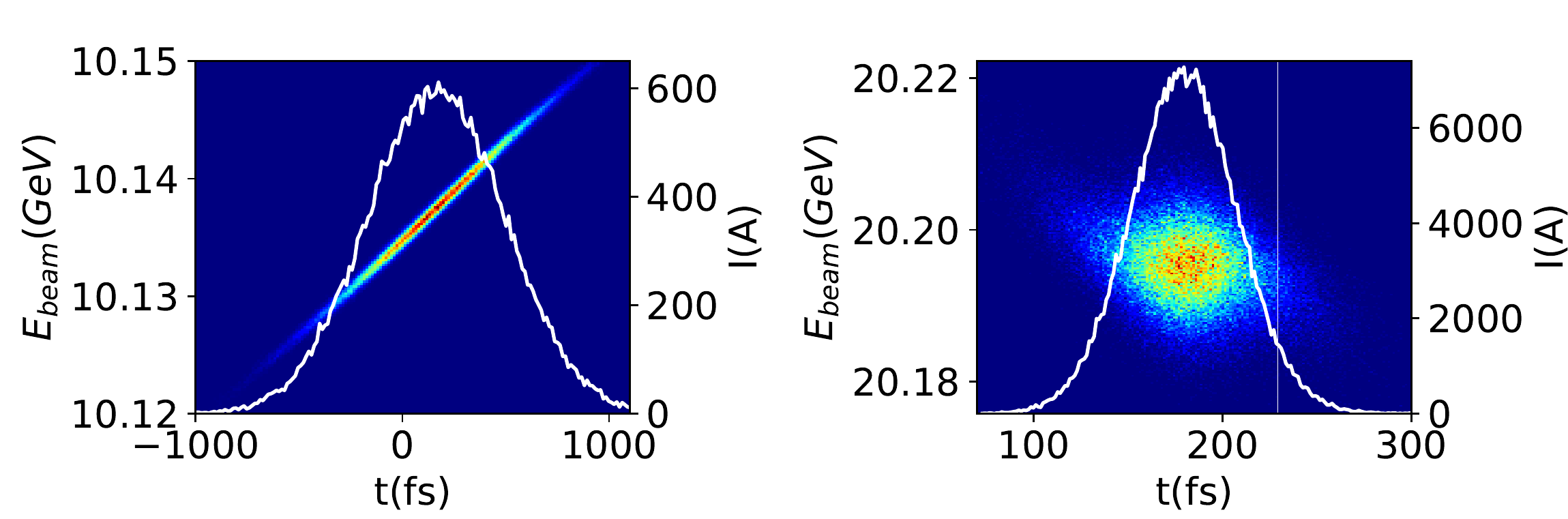} \caption{\label{fig:ps_20Gev} Beam distribution in longitudinal phase space
after passing through linac 1 (left) and linac 2 (right) for 20 GeV 
FEL operation, obtained
by tracking with ELEGANT \cite{elegant}, including the linac wake fields
from Ref.~\cite{Pellegrini:2235763}, 
and the shielded CSR impedance
from CSRZ \cite{zhoujjap}. The white dash-dotted line represents the
current profile. The final FWHM bunch length 
is 63 fs, or 19 $\mu$m, and the fitted rms bunch length $\sigma_{z}=8.5$~$\mu$m.}
\end{figure}

\begin{figure}[htbp]
\includegraphics[width=0.48\textwidth]{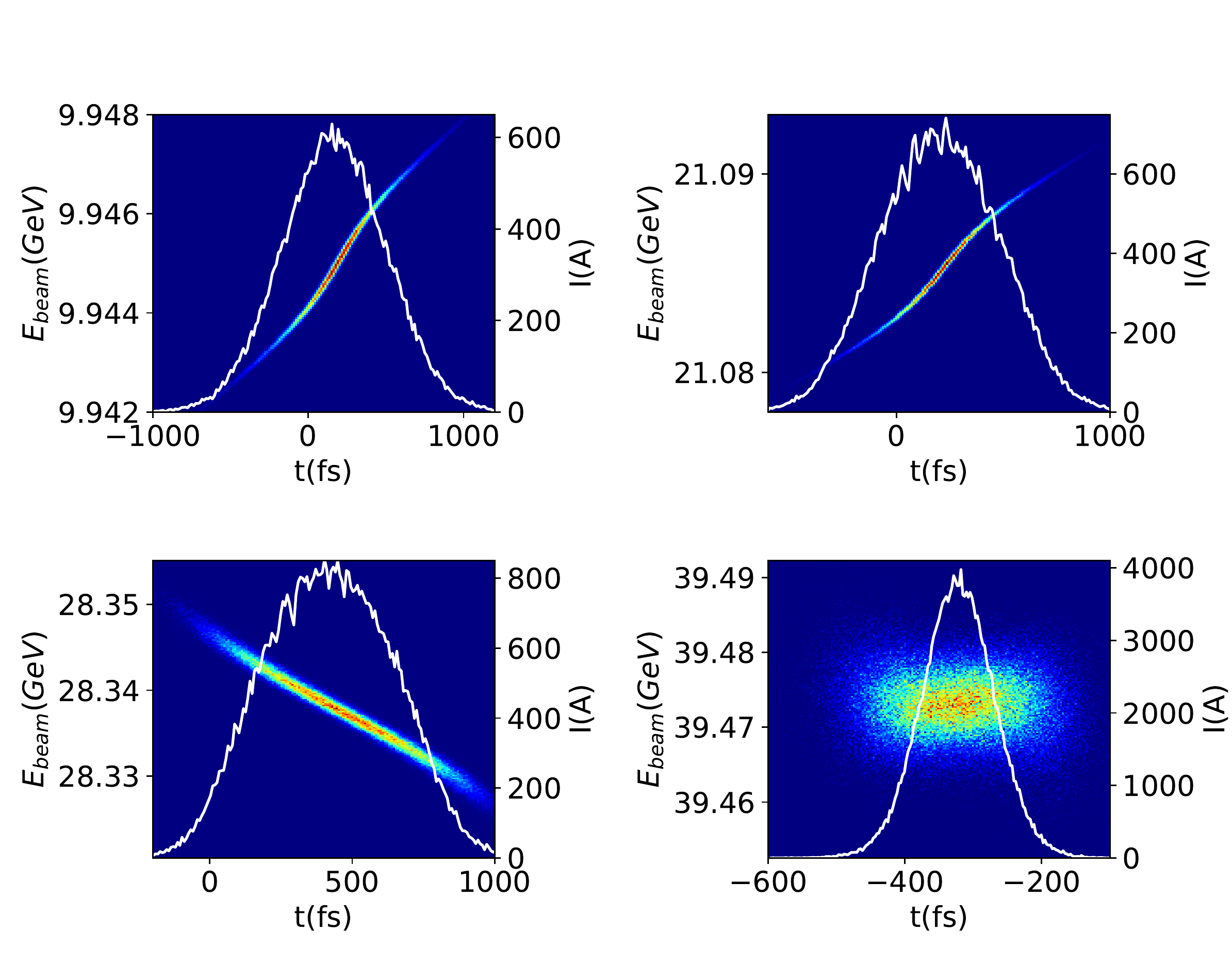} 
\caption{\label{fig:ps_40Gev} Beam distribution in longitudinal phase space 
after passing through linac 1 (top left) linac 2 (top right), linac 3 (bottom left) and linac 4 (bottom right)
for 40 GeV FEL operation, 
obtained by tracking 
with ELEGANT \cite{elegant}, including the linac wake fields from Ref.~\cite{Pellegrini:2235763}, 
and the shielded CSR impedance from CSRZ \cite{zhoujjap}.}
\end{figure}
%%%%%%%%%%%%%%%%

Figure~\ref{fig:ps_20Gev} shows the result of
the optimization for 20 GeV, obtained by tracking 100,000 particles in ELEGANT
through the first arc and two linac passages.
Figure \ref{fig:ps_40Gev} presents the result of
the optimization at 40 GeV, again 
obtained by tracking 100,000 particles in ELEGANT
now through three arcs and four linac passages. 
Tracking a larger number of 200,000 particles
yielded nearly identical results. 
Figures \ref{fig:ps_20Gev} and \ref{fig:ps_40Gev} show 
the simulated beam distribution in longitudinal phase after each of
the two or 
four linac passages, and superimpose the corresponding
bunch current profiles, a few parameters of which are compiled in
Table \ref{linacbp}.

\begin{table}[htbp]
\caption{Parameters characterizing the longitudinal bunch profile for each
linac passage: full-width-half-maximum (FWHM), FWHM divided by 2.355
(equal to a standard deviation $\sigma$ for a Gaussian profile), 
the rms bunch length $\sigma_{z}$ 
obtained from a Gaussian fit, and the peak current.
The numbers in parentheses refer to the 20 GeV case.
}
\label{linacbp} %
\begin{tabular}{lcccc}
\hline 
Linac  & FWHM  & FWHM/2.355   & fitted $\sigma_{z}$  & $\hat{I}_{b}$ 
\tabularnewline
 & [$\mu$m] & [$\mu$m]  & [$\mu$m] & [kA]
\tabularnewline
\hline 
linac 1  & 228  & 97  & 100  & 0.6\tabularnewline
\multirow{2}{*}{linac 2} & \multirow{1}{*}{204} & 86.6  & 82.5  & 0.7\tabularnewline
 & (19)  & (8.3)  & (8.5)  & (7.3)\tabularnewline
linac 3  & 175  & 75  & 66  & 0.8\tabularnewline
linac 4  & 35  & 15  & 16  & 4\tabularnewline
\hline 
\end{tabular}
\end{table}

Table~\ref{lhecparam} summarizes the optimized electron beam parameters
for LHeC FEL operation. The bunch compression using three linac passages and 
three arcs increases the peak bunch current by more than an order of
magnitude while preserving a reasonable transverse emittance and energy
spread suitable for FEL operation.

\begin{table}[htbp] 
\caption{\label{lhecparam} The main LHeC-ERL electron beam parameters. Peak
current, bunch length, and transverse emittance were obtained from
the tracking simulation. The numbers in parentheses
refer to the 20 GeV case.}
\begin{ruledtabular}
\begin{tabular}{lcr}
Parameters  & Unit  & Value\tabularnewline
\hline 
injection energy  & GeV  & 0.5\tabularnewline
final energy  & GeV  & 40 (20)\tabularnewline
electrons per bunch  &  & 3$\times10^{9}$\tabularnewline
initial FWHM bunch length  & $\mu$m  & 234\tabularnewline
final FWHM bunch length  & $\mu$m  & 35 (19)\tabularnewline
initial peak beam current  & kA  & 0.6\tabularnewline
final peak beam current  & kA  & 4 (7.2)\tabularnewline
final hor.~normalized emittance  & $\mu$m  & 0.9 (0.4)\tabularnewline
final vert.~normalized emittance  & $\mu$m  & 0.4\tabularnewline
bunch spacing  & ns  & 25\tabularnewline
final rms energy spread  & $\%$  & 0.01\tabularnewline
\end{tabular}
\end{ruledtabular}
\end{table}

At 20 GeV,  going through the first linac
and the first arc, the bunch can be compressed  
by about a factor of 12 at
the location of the undulator, from an initial rms length of 100
$\mu$m down 
to an rms length of about 8~$\mu$m; see Fig.~\ref{fig:ps_20Gev}
and Table~\ref{linacbp}. 
For a beam energy of 40 GeV, using three linac
passages followed by three arcs,  
we achieve a bunch compression by 
about a factor of 9, 
down to an rms length of about 15 $\mu$m, as is 
illustrated in Fig.~\ref{fig:ps_40Gev}.

 For comparison, at LCLS II the rms
bunch length can be varied between 0.6 and 52~$\mu$m, with a nominal
value of 8.3~$\mu$m \cite{LCLS2}, and the nominal rms bunch length
of the European X-FEL is 25~$\mu$m \cite{XFEL}.

The purpose of the present paper is to demonstrate the capacity of the LHeC-ERL for high gain FEL operation. Precise beam dynamics simulation require separate, additional work, in particular detailed studies of the strong compression of electron bunches in the presence of both CSR and resistive wall wake fields.  

Concerning the initial beam parameters, we note that  
for the 20 GeV simulations, where the compression is   accomplished in the first arc, 
we considered an initial rms relative energy spread 
at 500 MeV of about $10^{-3}$ ($\sim 0.5$~MeV), 
as was also assumed 
in the LHeC design report \cite[Section 7.3.3]{LHeCdesign}.  
This energy spread proved sufficient 
to suppress the microbunching.  
In the case of 40 GeV simulations, we observed that 
microbunching does not occur even for a 
ten times lower initial relative rms energy spread 
at 500 MeV of $10^{-4}$ (50 keV), since the incoherent synchrotron radiation in the second and third arc introduces a much larger rms spread of 0.4 
and 1.6 MeV, respectively.

In our simulations, 
we have not included the resistive wall wake field directly.
In the SLC arcs, with their compact 
aluminum vacuum chamber,  
not only the resistive wall, but also the 
wake fields of bellows 
and beam-position-monitors were significant \cite{slc}. 
To explore the sensitivity to wake fields in general,
we have increased the magnitude of the 
linac wake fields by up to a factor of 5.
Always readjusting the linac RF phases, after bunch compression, we obtained a 
similar bunch length and the same, or even slightly higher, peak current 
as for the nominal linac wake fields.   
We expect that the same would be true for 
other wake fields that 
induce a correlated energy variation, of similar magnitude,   
along the length of the bunch.  
Instead of wake fields, 
it is the (random) energy spread introduced 
by the incoherent 
synchrotron radiation in the arcs 
which ultimately limits the achievable bunch length.

%%%%%%%%%%%%%%%%%%%%%%%%%%
%%%%%%%%%%%%%%%%%%%%%%%%%%
\section{FEL Considerations}

In a free-electron laser, the active medium is a beam of relativistic electrons. The FEL interaction amplifies the undulator radiation in the forward direction, leading to an exponential growth of the radiation power along the length of the undulator. A self-amplified spontaneous emission (SASE) FEL does not require any optical cavity, nor any coherent seed, and it can operate in the X-ray regime. 
The wavelength of the radiation is given by the 
well-known formula 
\begin{equation}
\lambda= \frac{\lambda_{u}}{2 \gamma^{2}} 
\left( 1+\frac{K^{2}}{2}\right) \; , 
\label{undwave} 
\end{equation} 
where  $\lambda_{u}$ denotes the period length of a (planar) undulator,  
$\gamma$ the relativistic factor, proportional to the electron energy,
and  $K$ the undulator parameter \cite{wiedemann}.

The optimum matching of the electron beam to the light beam is achieved 
under the diffraction limit condition
\begin{equation}
\label{condition}
\varepsilon_{N}\leq \gamma \frac{\lambda}{4\pi} \; ,
\end{equation}
where $\varepsilon_{N}\equiv \gamma \varepsilon$ signifies the normalized emittance. However, it has been demonstrated that FELs can still operate, albeit with a reduced efficiency, even if the normalized emittance exceeds this optimum condition by a factor of four to five~\cite{SLSdesign}. Consequently, we expect that FEL light of wavelength around 0.5~$\textrm{\AA}$ 
can be produced by 40 GeV electrons with a normalized rms emittance of 0.9 $\mu$m.

The concrete goal of our LHeC ERL based FEL design is to generate
hard X-ray FEL radiation in the range between about 0.5 \AA$ $ and 8 \AA.
Following the second linac, we consider an FEL line featuring a
planar undulator with 39 mm period length, similar to the soft X-ray
undulator (SXU) line for LCLS II \cite{LCLSUn,LCLSUn2}. 
The minimum gap of this kind of undulator is 7.2 mm, with a  
magnetic field at the minimum gap of 1.5 T, and a resulting 
undulator parameter $K$ of 5.5.
The planar undulator is characterized in Table~\ref{unduparam}. 
The targeted wavelength range can be covered by varying the
electron beam energy from 10 to 40 GeV, in steps of 10 GeV, 
and changing the $K$ value by opening the undulator gap. 

\begin{table}%[htb]
    \caption{\label{unduparam}Parameters of the
    planar undulator considered. }
   \begin{ruledtabular}
\begin{tabular}{lr}
    parameter                       & value 	  \\\hline
    period length [mm]           & 39	   \\
    number of periods             & 85         \\
    minimum gap [mm]                 & 7.2	  \\
    undulator parameter $K$                            & 5.5    \\ 
   photon wavelength range [\AA{}]         & 0.5--7.6    \\ 
   \end{tabular}
    \end{ruledtabular}
\end{table}
%
%%%%%%
  
  With 40 GeV beam energy, tuning the undulator gap would actually also give us access to 
  wavelengths shorter than 0.5~\AA. 
  Besides, the LHeC ERL based FEL even offers opportunities to generate sub 10 pm FEL 
  radiation. For this purpose, a second FEL line hosting 
  a ``Delta'' undulator with 18 mm period \cite{DeltaU} can be employed. 
  However, obtaining and controlling the transverse coherence 
  for the shorter wavelengths would benefit  from a smaller 
  transverse emittance of the 40 GeV electron beam 
  (see Eq.~(\ref{condition})). 
  Results of a preliminary study for a sub-10-pm 
FEL line are reported 
  in Subsection~\ref{sec:Incoherent}.

%%%%%%%%%%%%%%%%%%%%%%%%%%%%%%%%%%%%%%%\
%%%%%%%%%%%%%%%%%%%%%%%%%%%%%%%%%%%%%%%
%%%%%%%%%%%%%%%%%%%%%%%%%%%%%%%%%%%%%%%
\subsection{\label{sec:FELperformance}FEL Performance}
\begin{figure}
\includegraphics[width=0.45\textwidth]{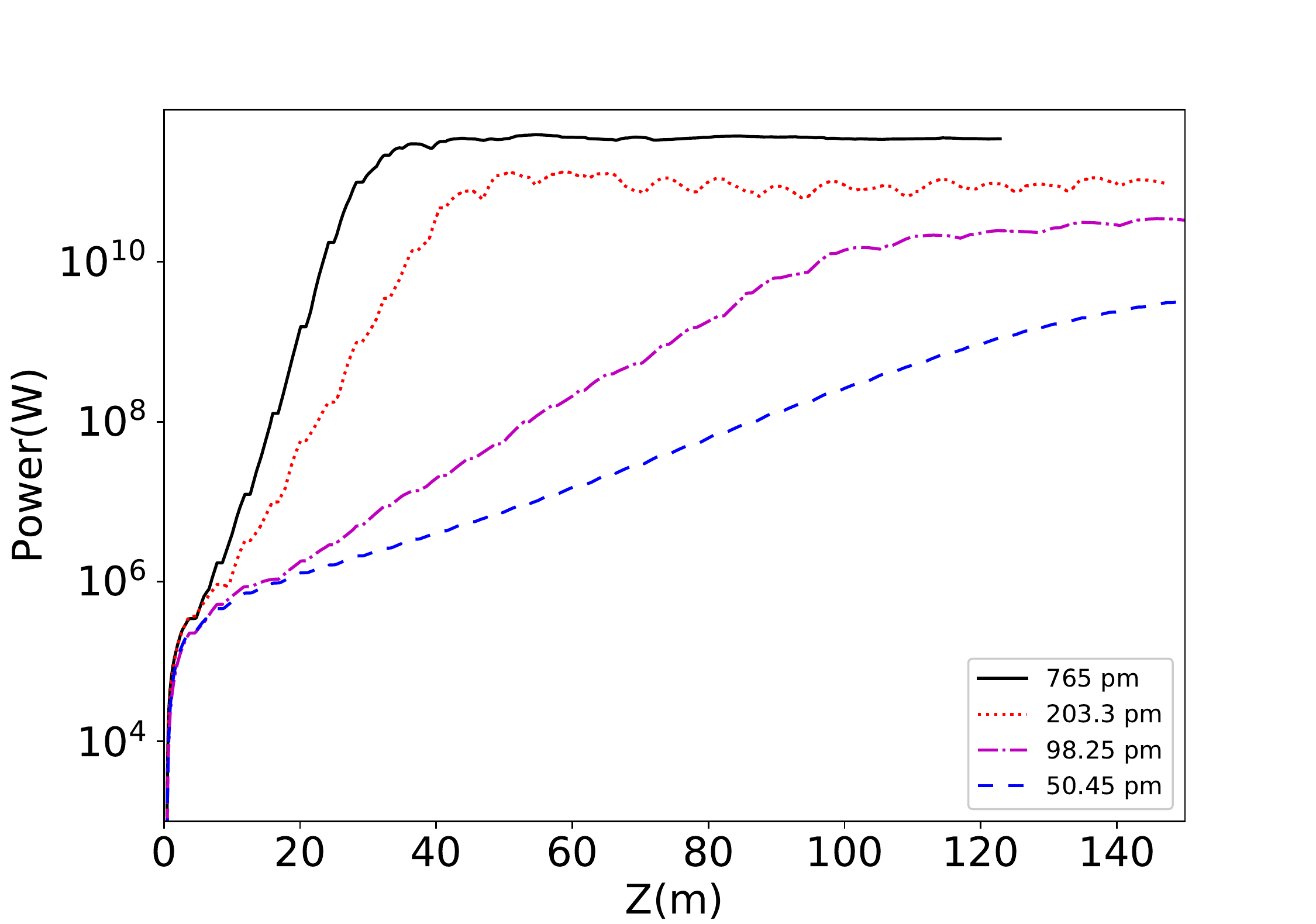} 
\caption{Growth of photon pulse power  at 
7.6 $\textrm{\AA}$  (black line)  2 $\textrm{\AA}$  (red dotted), 1 $\textrm{\AA}$ (magenta dot-dashed) 
and 0.5 $\textrm{\AA}$ (blue dashed) for an LHeC electron beam of energy
10, 20, 30 and 40 GeV, respectively,  
passing through the undulator FEL line with period 
$\lambda_u=39$~mm, as simulated with the code GENESIS. 
} \label{fig:FEL_powerG}
\end{figure}

\begin{figure*}[!hbt]
\includegraphics[width=0.9\textwidth]{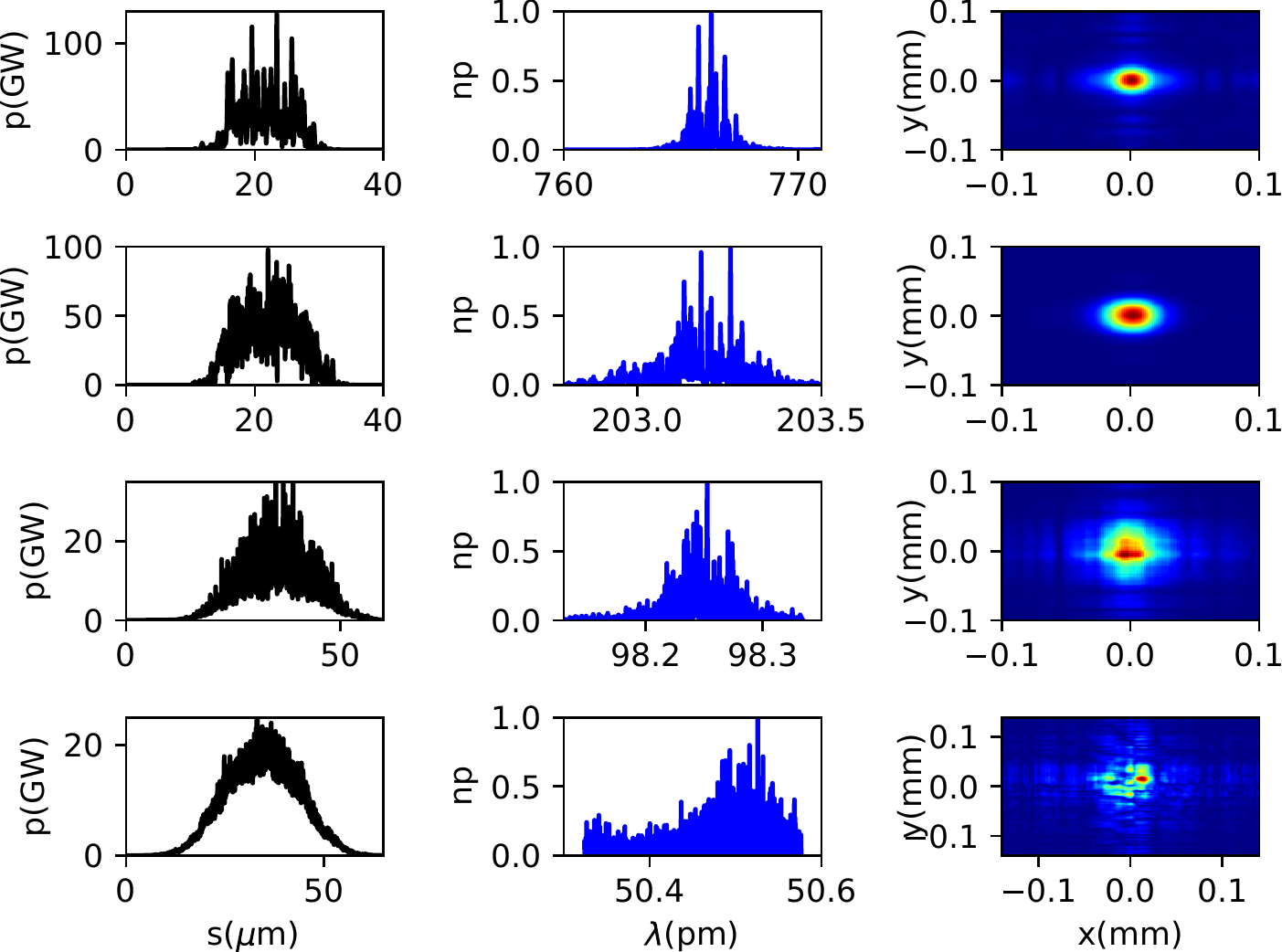} 
\caption{\label{fig:FELPulse} 
Spatial profile of the radiation pulse (left), 
wavelength spectrum of the radiation (centre) and 
transverse cross section of the of FEL radiation pulse around 
the point of saturation (right) for a beam energy of 10, 20, 30 and 40 GeV (from top to bottom), 
as simulated with GENESIS, using the respective distribution of the accelerated beam, obtained 
from ELEGANT, as input.  
}
\end{figure*}

Three-dimensional time-dependent simulations of the FEL process 
have been performed  with the  code GENESIS \cite{genesis}. The
electron beam energies considered ---  10, 20, 30 and 40 GeV --- correspond 
to photon wavelengths of about 7.6, 2.0, 1.0 and 0.5~$\textrm{\AA}$, respectively. 
For the undulator beam line, a FODO lattice,   
with a half cell length of 4.095~m,  
was selected for its simplicity and cost-effectiveness, 
since it limits the total number of additional magnets. 
The length of each undulator is 3.315~m. 
Undulator modules are separated by intervals of 780 mm, 
providing some space for focusing, steering, 
diagnostics or vacuum-system components. 
Figure~\ref{fig:FEL_powerG}
shows the simulated power growth at different FEL wavelengths 
generated by electron beams of the corresponding  energies. 
Depending on the wavelength the saturation occurs after a 
distance varying between 30 m and about 120 m.
Figure \ref{fig:FELPulse} presents the spatial profile of the radiation
pulses (first column), the spectrum of the radiation (second column), and
the transverse cross section of the FEL radiation around the
point of saturation, 
for beam energies of 10, 20, 30 and 40 GeV (from top to bottom). 
%%%%%%%%%%%%%%%%

\subsection{\label{sec:undulatorwake}Undulator Wake Field}
Longitudinal wake fields inside the undulator could increase the relative
energy spread within the bunch, which for efficient lasing must stay
less than a few times the Pierce parameter $\rho$ \cite{Bonifacio:1984qs,Huang:2006cq}.
The dominant wake field inside the undulator is due to the resistive
wall. Bane and Stupakov showed, for LCLS undulators, that taking into
account the ac conductivity, a flat aluminum chamber is preferred
over a round copper chamber, and that the anomalous skin effect can
be neglected \cite{Bane:2005kq}. In this case the peak of the wake
function has an amplitude of about $Z_{0}c/(a^{2}c)$ where $a$ denotes
the vertical half gap \cite{Bane:2005kq}, $Z_{0}$ the vacuum impedance
(about 377 $\Omega$), and $c$ the speed of light. The effect of
the wake field scales in first order with the bunch population $N_{b}$,
and with the inverse of the beam energy $E_{b}$. For short bunches
the wake field is independent of the bunch length, for long bunches
it scales with the inverse 3/2 power; the transition between the two
regimes occurs for bunch lengths of a few $\mu$m to tens of $\mu$m
\cite{Chao:1993zn}, depending on beam pipe radius and surface resistivity.

Compared with the LCLS, the LHeC FEL bunch lengths are roughly 
a factor 2 shorter (10~$\mu$m versus 20 $\mu$m), but the bunch charge of
the LHeC FEL is a factor 2 lower (0.5 nC versus 1 nC), the beam energy
up to a factor 3 higher (40 GeV vs.~14 GeV). Combining these factors,
for equal undulator length $L_{u}$ and beam pipe radius ($L_{u} \approx 130$~m, 
and $a=2.5~$mm for the LCLS \cite{Bane:2005kq}), the energy spread
induced by the undulator wake field for the LHeC FEL 
should be less important than for the LCLS.

In addition, the average energy loss due to wake fields, arising along the
length of undulator, could be partly compensated by tapering the field
strength of the undulator as a function of longitudinal location.

The resistive-wall wake field does not only affect the beam, 
but it also leads to a significant heat load 
inside the undulator, which will need to be considered 
in an engineering design for the LHeC ERL-FEL.

%%%%%%%%%%%%%%%%%%%%%%%%%%%%%%%%%%%%%%%%
%%%%%%%%%%%%%%%%%%%%%%%%%%%%%%%%%%%%%%%%

\subsection{\label{sec:brilliance}FEL Brilliance}
\begin{figure}[htb]
\includegraphics[width=0.48\textwidth]{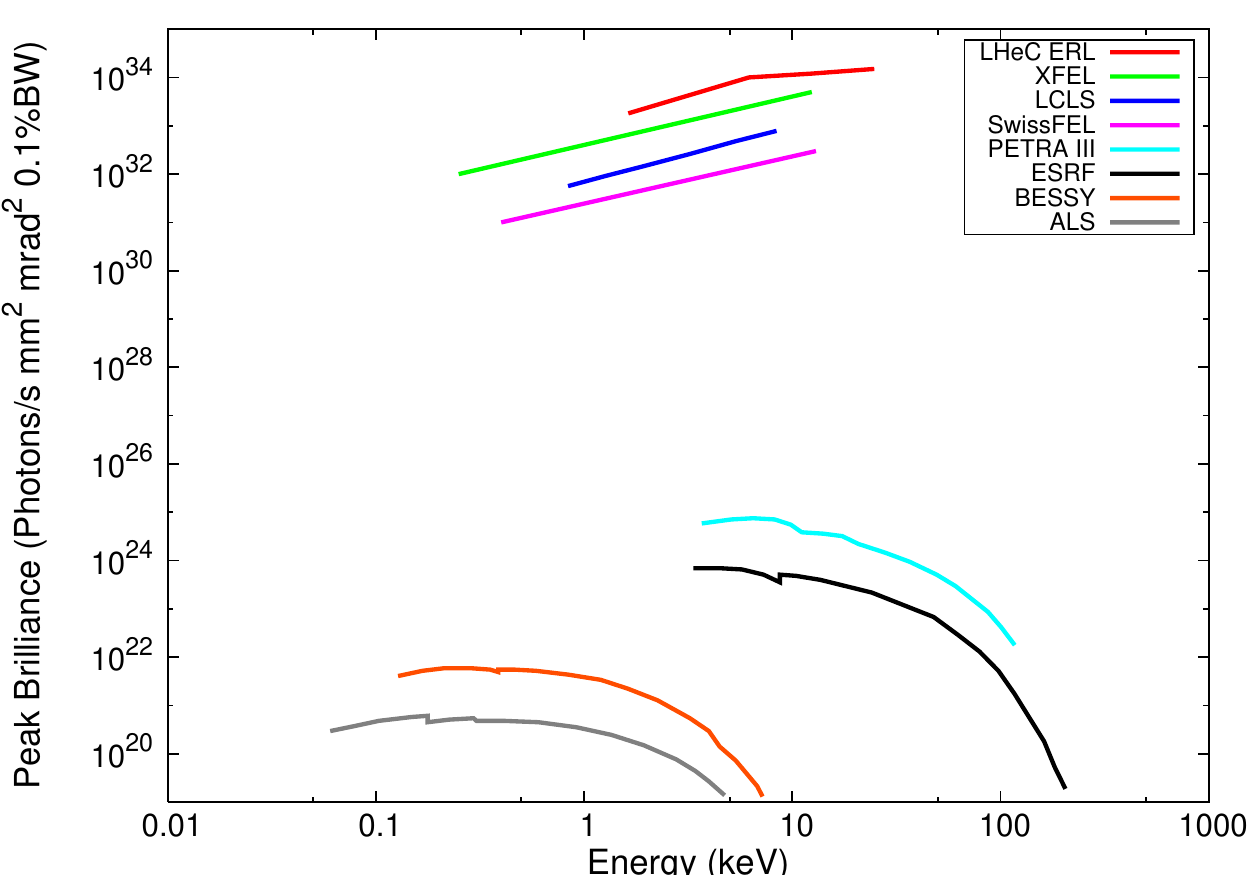}
\hspace{1 cm}
 \includegraphics[width=0.48\textwidth]{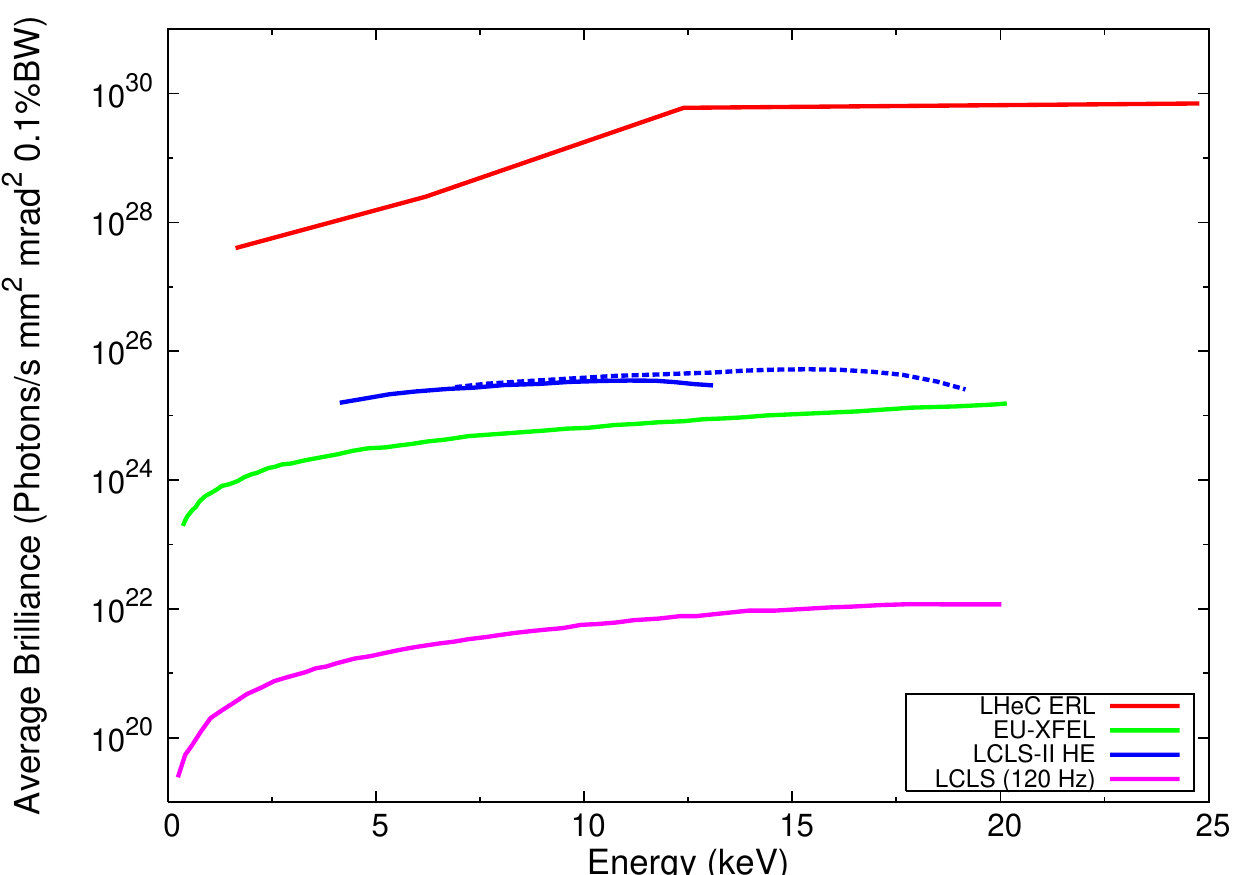}
\caption{\label{fig:bril}Comparison of FEL peak 
and average brilliance for the LHeC-FEL
with several existing or planned hard X-ray FEL and SR sources \cite{focus}.}
\end{figure}

One of the important parameters for comparing different radiation
sources is the brilliance \cite{saldin}. The brilliance describes
the intensity of a light source including its spectral purity and
opening angle. It can be calculated from the spectral flux (in units
of photons/s/$0.1\%$ bandwidth) by using the relation 
\begin{equation}
B=\frac{{\rm spectral\;flux}}{4\pi^{2}\Sigma_{x}\Sigma_{x}^{\prime}\Sigma_{y}\Sigma_{y}^{\prime}}\;,\label{brilliance}
\end{equation}
with the quantities 
\begin{equation}
\Sigma=\sqrt{\sigma_{e}^{2}+\sigma_{ph}^{2}}\label{sigma}
\end{equation}
and 
\begin{equation}
\Sigma^{\prime}=\sqrt{\sigma_{e}^{\prime2}+\sigma_{ph}^{\prime2}}\;.\label{sigmaprime} \; , 
\end{equation}
where $\sigma_{e}$, $\sigma_{e}^{\prime}$, $\sigma_{ph}$ and $\sigma_{ph}^{\prime}$
denote the transverse rms sizes and angular divergences of electron and photon
beams \cite{schmuser}. In the case of full transverse coherence
$\Sigma\Sigma^{\prime}=\lambda_{ph}/(4\pi)$. The brilliance values
for our four cases are listed in Table~\ref{FELparam}, along
with some other FEL parameters. A comparison of the LHeC ERL-FEL with
a few existing and planned hard X-ray sources 
\cite{LCLS2,brachmann,SLSdesign,XFEL,LCLS2_2}  
is presented in Fig.~\ref{fig:bril}.
These figures demonstrate 
that the peak brilliance of the LHeC ERL-FEL is as high as 
the one of the European XFEL, while the average 
brilliance is orders of magnitude higher, thanks to the high
average beam current, enabled by energy recovery.

The relatively high value of the horizontal emittance at 40 GeV causes a decrease in 
brilliance at wavelengths less than 1~\AA . 
We note that the estimate of the LHeC ERL-FEL brilliance in this region is approximate, 
as the radiation is no longer fully coherent. 

%%%

%%%%%%%%%%%
\begin{table}[htb]
    \caption{\label{FELparam} LHeC ERL-FEL radiation parameters 
derived from GENESIS simulations. 
The unit for the corresponding peak and average 
brilliance (B) is equal to photons/mm$^{2}$/mrad$^{2}$/s/$0.1\%$bw.}
\begin{ruledtabular}
    \begin{tabular}{lcccc}
     electron energy (GeV) &  10           & 20                &30       &40 \\\hline
    wavelength  (A)   & 7.6	      &  2.0           & 1     &0.50	    \\
    photon energy (keV)  & 1.63 	         &  6.2  	     &12.4       &24.8         \\
   saturation length (m) & 30                &  40	     & 100     & 120            \\
    peak power  (GW)     & 70         & 18           &   5             & 1.7      \\ 
    pulse duration (fs)  & 60     & 60   	      &  120             & 120         \\
   bandwidth ($\%$)      &  0.1          & 0.06       & 0.04        &  0.04   \\   
    photons per pulse ($\#\times 10^{10}$) & 1600      & 360     & 150              &  50   \\
    peak brilliance (B$\times 10^{32}$)   & 18	       &  100     &   120         &  150     \\
    average brilliance (B$\times 10^{27}$)   &  4	       & 25     &   65          &  70\\
    \end{tabular}
  \end{ruledtabular}
\end{table}
%%%%%%%%%%%%%%%%%%%%%

%

%%%%%%%%%%%%%%%%%%%%
Since the LHeC energy recovery linac provides a high-current, high-energy
and high repetition rate electron beam, the average brilliance of the LHeC-FEL is greater,
by at least three orders of magnitude, than for any other FEL source in operation
or under construction in the world. It also is about two orders of magnitude
higher than the projected average brightness predicted for ERL-extensions
of presently existing X-ray FEL infrastructures, as, e.g., 
in Ref.~\cite{sekutowicz}.  
Handling this bright a photon beam will be challenging; it is likely that during the commissioning of the proposed facility the average beam current can only be raised slowly, as various technical obstacles might be encountered and need to be addressed.

%%%%%%%%%%%%%%%%%%%%%%%%%%%%%%%%%%%%%%%%
%%\subsection{Incoherent sub 10 pm FEL radiation}

\subsection{\label{sec:Incoherent}Picometer FEL Radiation}

To set foot in the domain of even shorter wavelengths, that is the sub-10-pm region, we need to deploy an undulator with shorter period length. 
For this purpose, we consider a ``Delta undulator'' with 18 mm period and 5 mm minimum gap. 
The Delta undulator \cite{DeltaU,Bilderback_2010} is one 
of the best undulator sources for shaping the FEL 
photon polarization, and an example is 
currently employed at LCLS I. 
This type of undulator was originally 
proposed by A.~Temnykh, 
who designed, built and tested a prototype Delta undulator with 24 mm period at Cornell University \cite{DeltaU}. 
After this, Bilderback et al.~proposed a Delta undulator with 18 mm period and 5 mm minimum gap for an ERL-based coherent hard X-ray source \cite{Bilderback_2010}. 

\begin{figure}
\includegraphics[width=8cm]{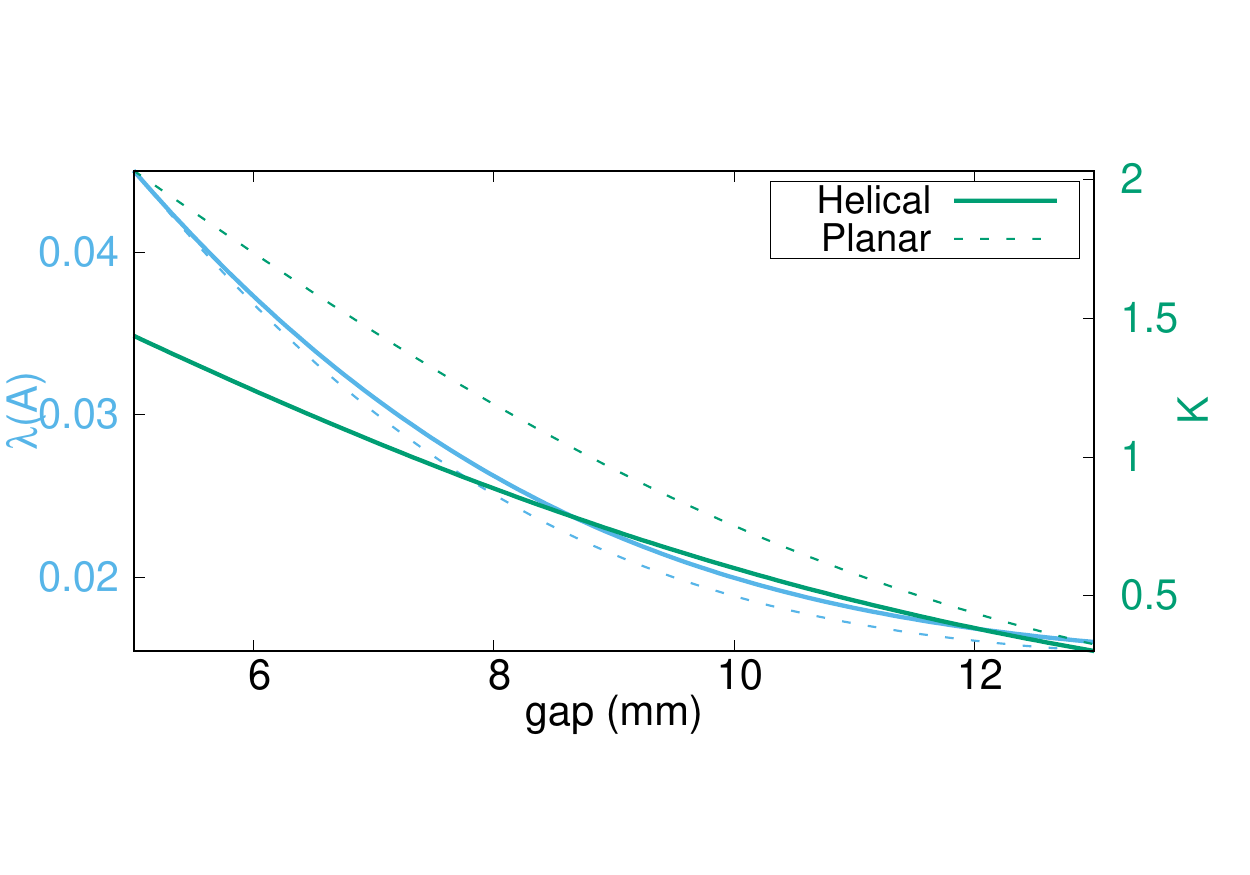}
\caption{\label{fig:DelatU} Radiation wavelength (left axis, blue)
and $K$ value (right axis, green) 
for a 40 GeV electron
beam passing through the Delta undulator as a function of 
the gap of undulator, in case of helical (solid lines) or planar
mode of operation (dashed lines). 
The period length of the undulator is taken to be 18 mm. }
\end{figure}

Our motivation for using this type of undulator source at the LHeC FEL 
is the prospect of producing radiation at wavelengths shorter 
than 0.07~\AA\ (7 pm) with a 30--40 GeV electron
beam. Figure \ref{fig:DelatU} illustrates the $K$ parameter 
and the radiation wavelength of a 40 GeV electron 
beam as a function of the gap size, for both planar and helical
operation mode of the Delta undulator,  as obtained by  
applying Eq.~(1) of Ref.~{\cite{DeltaU}}.  
Green (blue) dashed and solid lines show the $K$ value 
(linked to the radiation wavelength) 
for the helical and planar Delta undulator, respectively.
%%%%%%%%%%%%%%%%%%%%%%%%%%%%%%%%%%%%%%%% Figure Delta undulator

%%%%%%%%%%%%%%%%%%%%%%%%%%%%%%%%%%%%%%%%%%%%
\begin{figure}
\includegraphics[width=0.48\textwidth]{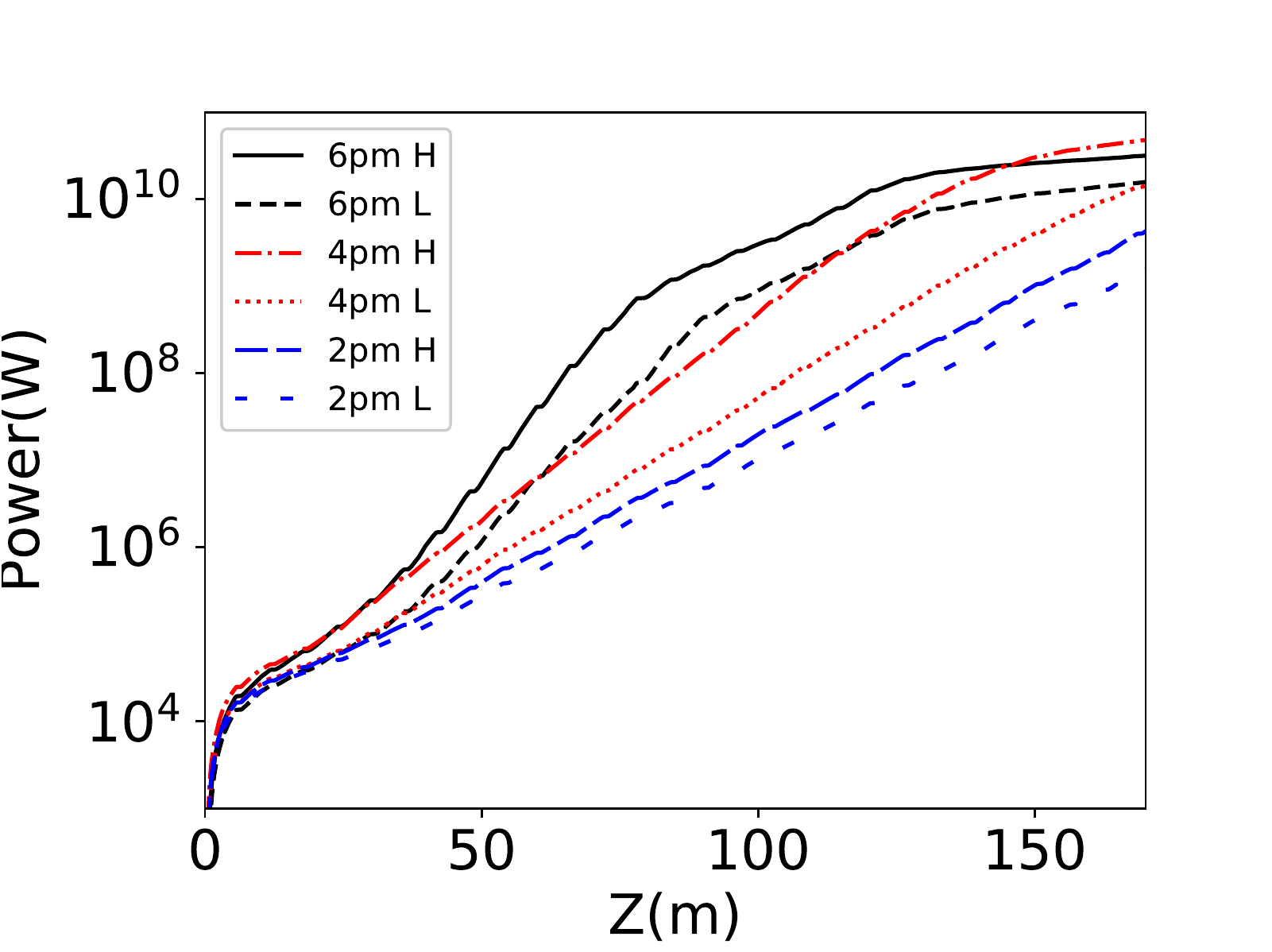} 
\caption{Simulated power growth for cases of helical (H) and linear (L) 
polarization of sub 10 pm radiation wavelengths. 
The simulations were performed for an electron beam of 
either 30 GeV (black lines) or 40 GeV 
(red and blue lines) passing through a helical or planar 
Delta undulator FEL line. \label{fig:power-growth40Gev}}
\end{figure}
%%%%%%%%%%%%%%%%%%%%%%%%%%%%%%%%%%%%%%%%%%%%

Figure \ref{fig:power-growth40Gev} presents our GENESIS 
simulations for the 
Delta undulator FEL line. The helical set up 
of the Delta undulator produces
helical polarization, the planar set up linear polarization. 
The simulations at 6 pm wavelength were performed for a 
30 GeV electron beam passing through the Delta undulator, 
with a gap of $\sim 6.5$ mm, considering 
either helical or linear polarization, 
shown by the black solid and dashed line, respectively. 
The figure also presents the growth at radiation wavelengths of 4 and 2 pm,
represented by the red and blue lines, 
for a 40 GeV beam passing 
through the helical or planar Delta undulator line.

Although both quantum fluctuations and slippage effects 
are included in these simulations, 
GENESIS simulations for wavelengths shorter than 10 pm 
may not be fully reliable. 
The reason is that GENESIS calculates the initial  
bunching factor from the number of macro-particles 
($N_{\rm part}$) found over the distance of
one wavelength, which, due to shot noise, would be  
$\left\langle b \right\rangle =1/\sqrt{N_{\rm {part}}}$, and calculates the 
radiation power from this bunching factor. 
At longer wavelengths, the
macro-particle number is usually less or equal to the actual number
of electrons in one ``beamlet'' (i.e., found over the distance of
one wavelength). For radiation
wavelengths shorter than 10 pm, and with 4 kA peak current, 
the actual number of electrons in one beamlet is 
only a few hundred electrons. 
If the number of macro-particles
inside a beamlet is lower, the GENESIS simulations
may not reveal the correct 
sensitivity to the transverse profile.
Conversely, if this number of macro-particles
is higher than the actual number of electrons in a beamlet,
the shot noise and bunching factor 
will be lower than in reality. 

In view of these considerations,  
and to validate our simulation results, we have 
benchmarked them against estimates 
from 1D and 3D FEL theory.
 The FEL gain length in 1D is $L_{G0}=\lambda_w/(4\pi \rho_{\rm 1D})$,
with  $\rho_{\rm 1D} \approx 1.4\times 10^{-4}$ denoting the 1D 
FEL parameter \cite{Saldin_2010}, 
evaluated for a wavelength of 4 pm. 
Therefore, the one dimensional gain length for the 
helical Delta undulator is around 6 m,
and the FEL power in saturation 
($P_{\rm sat}=\gamma mc^2 I/\rho_{\rm 1D}$) is  approximately 22 GW.

Taking into account the 3D effect on the FEL performance 
according to the methodology of E.~Saldin et al.~\cite[Eqs.~(3)--(5)]{Saldin2004}, 
the 3D gain length at 4 pm wavelength increases to around 
$L^{E.S.}_{G,3D}=17$ m. The gain length in our simulation 
is almost 20 m  (see Fig.~\ref{fig:power-growth40Gev}). 
Accordingly, the results of our GENESIS simulations 
are not far from the 3D FEL theory of Ref.~\cite{Saldin2004}. 
We can also consider another analytical model for the 3-D FEL effect, 
namely the one of M.~Xie \cite{Xie}. 
According to Xie's analysis, assuming an electron beam well
 matched to the design optics, the
3D power gain length, as a function of the average betatron 
function in the undulator, is calculated as 
$L^{M.X.} _{G,3D}=L_G(1+\Lambda),$ where $\Lambda$ includes 
the effects of the radiation diffraction, the electron beam 
transverse emittance and the uncorrelated energy spread \cite{Xie}. 
By using the values for our beam, 
the Xie formalism predicts $\sim 22$~m gain length, 
which is again quite close to our simulation result. 
From these comparisons, 
we conclude that the GENESIS simulation results for 
wavelengths of a few pm, presented  
in Fig.~\ref{fig:power-growth40Gev},  
are in good agreement with FEL theory.

%%%%%%%%%%%%%%%%%%%%%

Another issue of potential concern is that, in simulations with unprecedentedly 
high values of the photon energy, the
recoil effect on the emitting beam particle may become important.
At the wavelengths where this happens 
GENESIS will no longer produce correct results. 
The importance of the recoil is indicated by  
the quantum FEL parameter $\bar{\rho}$, defined as $\bar{\rho}=\rho \gamma mc/(\hbar k)$,
which represents the ratio between the 
classical maximum induced momentum spread and the 
one-photon recoil momentum \cite{bonifacio,quanFEL}. 
% The FEL dynamics depends on $\bar{\rho}$: 
If $\bar{\rho}\leq 1$, the FEL will exhibit a strong quantum recoil effect.  
Calculating the quantum FEL parameter 
at 2 pm wavelength (considering the helical Delta undulator with $K$=0.65), 
we find $\bar{\rho}\approx 5.9$, which is 
larger than 1. 
The resulting quantum recoil parameter 
$1/\bar{\rho}$ is 0.17. 
These numbers indicate that even at a wavelength of 2 pm the LHeC FEL dynamics remains essentially 
classical and is not strongly  
altered by the quantum recoil momentum. 
We note that, although the photon energy is high, 
the beam energy is much higher still, which explains the weak 
quantum recoil effect despite the short wavelength. 
In this case, the value of  
$\bar{\rho}$ indicates the number of 
resonant photons emitted per electron at saturation \cite{quanFEL}. 

In consequence,
the simulations of Fig.~\ref{fig:power-growth40Gev} 
inspire confidence that 
the LHeC FEL can produce more than 1 GW FEL peak power at wavelengths shorter than
10 pm. 
This mode of operation in the pm wavelength regime
could be another outstanding feature of the proposed new facility.
We can even consider the higher harmonics of these few pm radiation lines. 
Specifically, it is well known that the higher harmonics of
the radiation in helical undulators contain higher orders of the angular
momentum $\ell=(h-1)$ \cite{OAM1,OAM2}, where $h$ 
denotes the number of the harmonic.
Certainly, this ability can open a new pathway for studies of 
nuclear interactions.

In future studies of the short wavelength FEL operation 
based on the LHeC-ERL
we may investigate various possibilities to further enhance  
the efficiency of this facility in the few pm wavelength regime, 
and to advance the FEL performance for wavelengths
shorter than 50 pm, with the particular aim of improving the transverse coherence.
One idea would be to reduce the electron bunch charge, 
so as to be able to inject a beam with lower initial emittance, and, in addition, 
to better control the transverse emittance growth due to synchrotron radiation
by further optimizing the optics in the ERL arcs.
% quantum fluctuations in the ERL. 
% By improving the beam injection also we can reduce the emittance. 

%%%%%%%%%%%%%%%%%%%%%%%%%%%%%
%%%%%%%%%%%%%%%%%%%%%%%%%%%%%
%%%%%%%%%%%%%%%%%%%%%%%%%%%%%
% RECOVERY
%%%%%%%%%%%%%%%%%%%%%%%%%%%%%
%%%%%%%%%%%%%%%%%%%%%%%%%%%%%
%%%%%%%%%%%%%%%%%%%%%%%%%%%%%
\section{Energy Recovery}
\begin{figure}[!htb]
\centering \includegraphics[width=8cm]{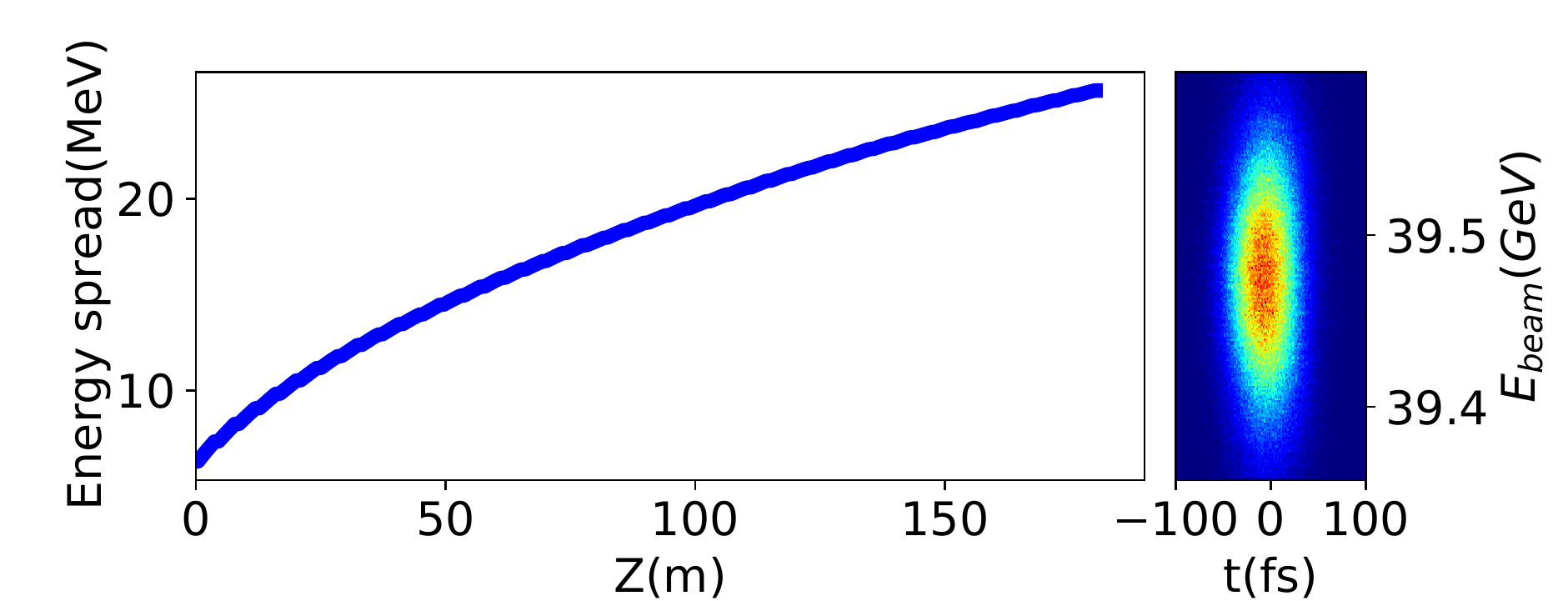}
\caption{(Left) The evolution of beam energy spread ($\sigma_{E}$) along the undulator
region for 0.5 \AA\ via 40 GeV e-beam energy through planar undulator. (Right) Longitudinal phase-space of e-beam after FEL radiation at 0.5~\AA . }
\label{fig:multiees} 
\end{figure}

The high average brilliance is achieved thanks to the high average
beam current, which relies on energy recovery.
For the energy recovery process, the energy spread
of the electron beam after the lasing process is an important parameter.
The evolution of this parameter is shown in Fig.~\ref{fig:multiees}
for an FEL wavelength of 0.5~\AA . 
Along the undulator, the relative 
energy spread increases approximately six times (from
0.01$\%$ to 0.06$\%$), but it remains small compared with
the energy acceptance of the optics. 
The energy spread at the saturation point ($z\approx$~120--150~m) 
is approximately 25 MeV. This value is low compared with the electron
beam energy, and also with the electron injection energy of 500 MeV.
It can further be reduced by energy compression    
% phasing of the downstream linacs. 
% and could also be partially corrected
in the downstream arcs and linacs.

\begin{figure*}[!htb]
\centering \includegraphics[width=1\textwidth]{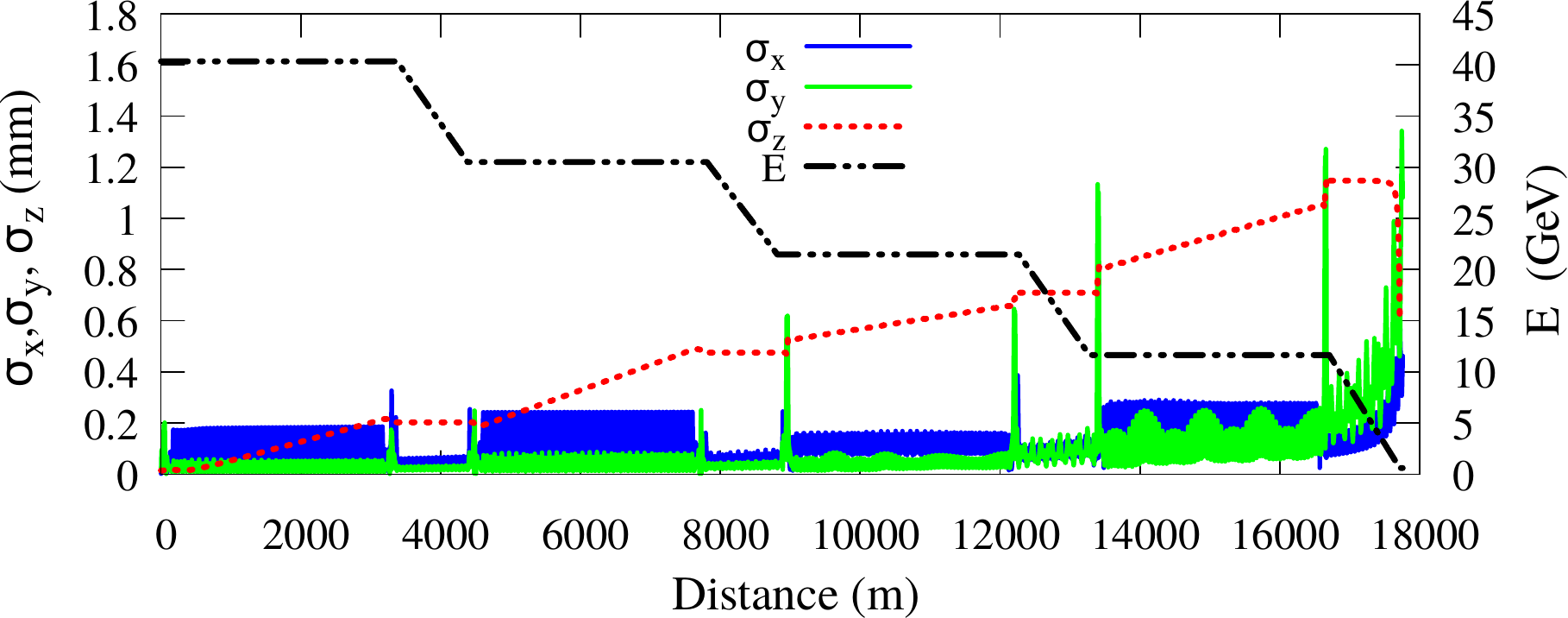} 
\caption{Beam energy and beta functions for the deceleration of the spent beam, after lasing
at 0.5~\AA ,  
over two complete LHeC turns starting from 40 GeV.}
\label{fig:twiss} 
\end{figure*}

To study this aspect further and to demonstrate the feasibility of
energy recovery during FEL operation, we have simulated the deceleration
process from the maximum beam energy about 40 GeV 
down to about 0.5
GeV, starting with the beam distribution exiting the undulator, shown in Fig.~\ref{fig:multiees}.
This distribution, modelled by 
$8\times10^{5}$ macroparticles representing a single bunch, was obtained from
the GENESIS FEL simulation for the 0.5~\AA\  case. 
We next used again the
simulation code ELEGANT 
%  PLACET \cite{placet2,placet}
 to track the $3\times10^{5}$ 
macroparticles through the exact optics \cite{LHeCdesign,bogaczarc}
for the last two decelerating turns (four arcs and four linac passages)
of the LHeC, composed of 16,000 beam-line elements. 
As before for the acceleration, also here both 
the linac wake fields and the shielded CSR in the arcs were 
taken into account. To control energy
spread and bunch length during deceleration the bunch arrival phase
in the linacs was set to $-170^{\circ}$ instead of the $-180^{\circ}$
which would correspond to maximum deceleration. Figure \ref{fig:twiss}
shows the simulated beam size, bunch length and beam energy during the 
deceleration process. In the simulation,  
not a single macroparticle was lost. 
The final rms beam of order 1 mm, 
is much smaller than   
the linac RF cavity iris radius of 7 cm \cite{Calaga:2020926}.  
We have verified that deceleration is 
also possible, and even 
easier, for the 20 GeV 
single-turn ERL operation.

%\

\section{Applications for an Extremely Brilliant Coherent X-Ray Source}
The  brilliant photon beams 
at wavelengths below 1~\AA\ 
generated by the LHeC FEL 
could potentially revolutionize
scientific experiments in different fields of research such as 
biology, chemistry, material science,
atomic physics, nuclear physics, and
particle physics.

High-resolution high-brilliance X-rays, with wavelengths of less than
1~\AA\  would allow advanced imaging of enzymes \cite{VRIELINK2003709},
viral assemblies \cite{phduc}, and corona viruses \cite{Hilgenfeld},
and, e.g., enable more efficient antiviral drug design \cite{Hilgenfeld}.
Shorter wavelength dramatically improves atomic resolution data (e.g.~approximately
five times more data are expected 
to be available at 0.95~{\AA} resolution
than at 1.5~{\AA} resolution \cite{VRIELINK2003709}).

Hard X-rays with photon energies exceeding 10 keV 
($\lambda<1.2$~\AA ) 
also enable studies of thick 3D materials due to their deep penetration 
paired with excellent spatial resolution. Such X-ray radiation
allows probing condensed matter systems on the atomic length scale
with minimum unwanted absorption. 

One of the possible applications of LHeC FEL would be resonant inelastic
X-ray scattering (RIXS) experiments. RIXS offers the unique capability
to record excitation spectra from complex materials by measuring the
momentum and energy dependence of inelastically scattered photons~\cite{stfc_}.
The cross section for RIXS scattering is extremely small compared
with other techniques such as elastic X-ray scattering or X-ray emission
spectroscopy. Therefore, the RIXS experiments require a high average
brilliance~\cite{RXIS_kim}.

Other ``photon-hungry'' experiments, which would be enabled by the
LHeC/ERL-based FEL include total X-ray scattering, X-ray diffraction
under high pressure, and resonant X-ray emission spectroscopy (RXES)
~\cite{xdsLima}. RXES is a powerful method for studying the electronic
structure of atoms, molecules and solid materials. The RXES signals
are much weaker than those of X-ray absorption spectroscopy (XAS),
so that, similar to RIXS, also RXES precision experiments require
a high-brilliance X-ray source \cite{marolt}.

As a 
concrete example, studies of nano-materials for advanced battery technologies
could greatly benefit from the high average brilliance available at
the LHeC-FEL~\cite{heminger}.

In general, the high average brilliance of the LHeC-FEL will facilitate
the detection of ultrafast changes of structures and of the electronic
states of natural and artificial materials~\cite{yabashi}. 

In addition, the proposed picometer FEL may prove 
a unique source of high-energy photons 
carrying orbital angular momentum,
as an alternative 
to the proposed inverse Compton scattering 
of twisted laser photons off a 
relativistic electron beam
\cite{nguyenprl}.

Finally, in the area of 
particle physics, the 
unique average intensity and the 
wide photon-energy range 
of the LHeC FEL radiation 
could enable intriguing  
hunts for  New Physics \cite{nguyen},  
including searches for Dark 
Photons and Axion-like Particles (ALPs)
\cite{nguyen1,ferber}.

%%%%%%%%%%%%%%%%%%%%%%%%%%%%% Finaly :D :) 
%%%%%%%
\section{\label{sec:concl}Conclusions}

We have investigated the potential 
radiation properties of a SASE
FEL based on an 
Energy Recovery linac, such as the LHeC.  
% The LHeC electron beam is cw 
% with 25 ns bunch spacing, and
% has an energy of tens of GeV.  
Our simulations of the FEL process, for 
LHeC electron beams of 10, 20, 30 
and 40 GeV passing 
through a planar 
LCLS-II type undulator with 39 mm period, 
suggest that
FEL radiation in the few {\AA}ngstrom or sub-{\AA}ngstrom wavelength regime can be produced, at significant power and brilliance 
(see 
% Fig.~\ref{fig:avbril} and 
Table \ref{FELparam}). 
Indeed, the LHeC-FEL promises 
an average brilliance far
exceeding those of other, existing or proposed X-ray FELs.

In addition to using 
a high-energy, cw electron beam with 25 ns bunch spacing, 
the high average brilliance relies on the following two features.
First, coherent synchrotron radiation is expected to be almost completely suppressed 
by realistic 
vacuum-chamber shielding, thanks to the large bending radius and
small vacuum chamber of the LHeC machine. 
This assumption has been validated 
by detailed simulations using the codes CSRZ and ELEGANT.
We note that these simulations
did not take into account any resistive-wall wake fields,
the magnitude of which was only estimated analytically.  
Second, we have shown that
the beam exiting the undulator can be decelerated efficiently from
40 GeV down to a few 100 MeV, without any noticeable beam loss, which is
the key prerequisite for the energy recovery mode of FEL operation.

The reported simulation results were obtained for the SASE FEL mode and without any tapering. By using self seeding and a tapered
undulator the performance could be further improved and the spectrum
be rendered more monochromatic.  
Furthermore, in combination with a low-loss crystal cavity, 
a free-electron laser oscillator 
operating in the {\AA}ngstrom wavelength regime could be realized  
\cite{PhysRevLett.100.244802}. 

We have also performed exploratory 
studies with a Delta undulator of 18 mm period, 
that could allow 
access to the extremely short 
wavelength range below 10 pm, using
the 40 GeV electron beam of the LHeC.

In summary, an ERL-based high-energy 
SASE FEL boasts various unique characteristics and offers 
tantalizing opportunities. 
The advent of such a facility would impact numerous 
areas of fundamental and applied science.

\section*{Acknowledgments}
We thank Herwig Schopper, the Chair of the LHeC International Advisory
Committee, for hinting at the use of the LHeC as an FEL. Particular
thanks go to Sven Reiche from PSI, and to 
Gianluca Geloni and Svitozar Serkez from the European XFEL,  
for enlightnening discussions on GENESIS simulations and effects at very short wavelengths.
We are grateful to
Alex Bogacz from Jefferson Lab for providing the LHeC
optics files in MAD-X \cite{madx} format, which we converted to
ELEGANT \cite{elegant}.
% , to PLACET \cite{placet2,placet}.  
We also
acknowledge continued encouragement from Oliver Br\"{u}ning and Max Klein.

This work was supported by the Turkish Atomic Energy Agency with Grant No.~2015 TAEK (CERN) A5.H6.F2-13, and by the European Commission under the HORIZON2020 Integrating Activity project ARIES, grant agreement 730871.

 \nocite{*}
\bibliography{main}

\end{document}